\documentclass{jfm}
\usepackage{graphicx}
\usepackage{epstopdf, epsfig}
\usepackage{amsmath}
\usepackage{xcolor}
\usepackage{tikz}
\newcommand\rgm[1]{#1}

\newcommand{\mylab}[3]{\raisebox{#2}[0mm][0mm]{\makebox[0mm][l]{\hspace*{#1}#3}}}

\newcommand{\blueline}{\raisebox{2pt}{\tikz{\draw[-,blue,solid,line width = 1pt](0,0) -- (6mm,0);}}}
\newcommand{\bluedashedline}{\raisebox{2pt}{\tikz{\draw[-,blue,dashed,line width = 1pt](0,0) -- (6mm,0);}}}
\newcommand{\reddashedline}{\raisebox{2pt}{\tikz{\draw[-,red,dashed,line width = 1pt](0,0) -- (6mm,0);}}}
\newcommand{\blackdashedline}{\raisebox{2pt}{\tikz{\draw[-,black,dashed,line width = 1pt](0,0) -- (6mm,0);}}}
\newcommand{\blackline}{\raisebox{2pt}{\tikz{\draw[-,black,solid,line width = 1pt](0,0) -- (6mm,0);}}}
\newcommand{\thicblackline}{\raisebox{2pt}{\tikz{\draw[-,black,solid,line width = 1.5pt](0,0) -- (6mm,0);}}}
\newcommand{\redline}{\raisebox{2pt}{\tikz{\draw[-,red,solid,line width = 1pt](0,0) -- (6mm,0);}}}
\newcommand{\lightblueline}{\raisebox{2pt}{\tikz{\draw[-,Colour4,solid,line width = 1pt](0,0) -- (6mm,0);}}}
\newcommand{\yellowline}{\raisebox{2pt}{\tikz{\draw[-,Colour7,solid,line width = 1pt](0,0) -- (6mm,0);}}}
\newcommand{\greenline}{\raisebox{2pt}{\tikz{\draw[-,darkgreen,solid,line width = 1pt](0,0) -- (6mm,0);}}}
\newcommand{\magentaline}{\raisebox{2pt}{\tikz{\draw[-,magenta,solid,line width = 1pt](0,0) -- (6mm,0);}}}
\definecolor{Colour1}{RGB}{   0,     0,255}
\definecolor{Colour2}{RGB}{255,    0,    0}
\definecolor{Colour4}{RGB}{  25.5,179,    255}
\definecolor{Colour7}{RGB}{237, 177, 32}
\definecolor{Black}{RGB}{0,0,0}
\definecolor{darkgreen}{RGB}{0, 127.5, 0}
\definecolor{magenta}{RGB}{255, 0, 255}

\renewcommand\d{\mathrm{d}}

\shorttitle{Turbulence and drag from texture-less simulations}
\shortauthor{W. Xie, C. T. Fairhall and  R. Garc\'ia-Mayoral}

\title{Resolving turbulence and drag over textured surfaces using texture-less simulations:\\ the case of slip/no-slip textures}

\author{  W. Xie \aff{1},
  C. T. Fairhall \aff{1}
  \and 
  R. Garc\'ia-Mayoral \aff{1}
  \corresp{\email{r.gmayoral@eng.cam.ac.uk}}
 }

\affiliation{\aff{1}Department of Engineering, University of Cambridge, Cambridge, CB2 1PZ, UK}
\begin{document}

\maketitle

\begin{abstract}
\noindent We study the effect of surface texture on an overlying turbulent flow for the case of
textures made of an alternating slip/no-slip pattern, a common model for superhydrophobic surfaces,
but also a particularly simple form of texture.
\rgm{For texture sizes $L^+ \gtrsim 25$, we have previously reported that, even though the texture effectively imposes homogeneous slip boundary conditions on the
overlying, background turbulence, this is not its sole effect. The effective conditions only produce
an origin offset on the background turbulence, which remains otherwise smooth-wall-like.
For actual textures, however, as their size increases from $L^+ \gtrsim 25$ the flow progressively
departs from this smooth-wall-like regime, resulting in additional shear Reynolds stress and increased drag, in a non-homogeneous fashion that could not be reproduced
by the effective boundary conditions. This paper focuses on the underlying physical mechanism
of this phenomenon. We argue that it is caused by the non-linear interaction of the texture-coherent
flow, directly induced by the surface topology, and the background turbulence,
as it acts directly on the latter and alters it. This does not occur at
the boundary where effective conditions are imposed, but within the overlying flow
itself, where the interaction acts as a forcing on the governing equations
of the background turbulence, and takes the form of cross-advective terms between the 
latter and the texture-coherent flow.}
%This flow component was, incidentally, responsible for %
%For texture sizes $L^+ \lesssim 20$, we have
%previously reported that turbulence remained smooth-wall-like, other than experiencing an apparent
%origin offset for different flow components. For slip/no-slip textures, this effect reduced to the
%flow experiencing slip conditions in the streamwise and spanwise directions and zero transpiration
%at the surface. The overlying turbulence effectively perceived such boundary
%conditions at least up to texture sizes $L^+ \approx 50$. However, beyond $L^+ \approx 20$ the
%texture interacted with the overlying turbulence in a non-homogeneous fashion, additional Reynolds
%stresses arose and turbulence was no longer smooth-wall-like. This is the typical effect of surface
%texture observed for rough surfaces, and results in an  increase in drag relative to smooth-wall flows.
%In this paper,
%we argue that this occurs because the texture modifies the overlying turbulence through
%non-linear, cross-advective terms between the background turbulence and the texture-coherent flow directly induced
%by the surface topology.
\rgm{We show this by conducting simulations where we remove the texture and introduce
additional, forcing terms in the Navier-Stokes equations, in addition
to the equivalent homogeneous slip boundary conditions. The forcing terms capture the effect of the non-linear
interaction on the background turbulence
without the need to resolve the surface texture. 
We show that, when the forcing terms are derived 
accounting for the amplitude modulation of the texture-coherent flow by the background turbulence, they
quantitatively capture the
changes in the flow up to texture sizes $L^+ \approx 70$--$100$. This includes not just the roughness function
but also the changes in the flow statistics and structure.
}
\end{abstract}


\section{Introduction}
\label{sec:intro}
This paper focuses on the effect on wall turbulence of surface texture, which alters the flow
and modifies the drag compared to a smooth surface. The most common example of surface texture
is roughness \citep{schlichting1937experimental, colebrook1937experiments, jimenez2004turbulent,
chung2021predicting}, which generally increases drag, but there are also textures
such as riblets \citep{walsh1984}, superhydrophobic surfaces \citep{Rothstein10} and anisotropic
permeable substrates \citep{abderrahaman2017analysis, gomez2019turbulent} that can reduce drag. 

According to the classical theory of wall turbulence, for small textures the only effect of the
surface on the outer flow is a uniform shift in the mean velocity profile, $\Delta U^+$, compared
to a smooth surface \citep{clauser1956turbulent,spalart2011drag,rgm2019}. The superscript
`+' indicates scaling in viscous units, i.e. normalisation by the the kinematic viscosity $\nu$
and friction velocity $u_{\tau} = \sqrt{\tau_w}$, where $\tau_w$ is the tangential stress at the
wall. Throughout the paper, we consider upward shifts of the velocity profile, which occur for
drag-reducing surfaces, positive , $\Delta U^+ > 0$, and downward shifts of the velocity profile,
which occur for drag-increasing surfaces, negative, $\Delta U^+ < 0$, so our $\Delta U^+$ is the opposite of the usual `roughness function' used for drag-increasing textures. The free-stream velocity, $U_{\delta}^+$, is then given by
\begin{equation}
U_{\delta}^+ = \left( \frac{2}{c_f} \right)^{1/2} = \frac{1}{\kappa}\log \delta^+ + B + \Delta U^+,
\label{eq:ud+}
\end{equation}
where $c_f$ is the skin friction coefficient based on $U_{\delta}$, $\delta$ is the flow thickness,
and the von K\'{a}rm\'{a}n constant, $\kappa$, and the parameter $B$, which includes the smooth-wall
logarithmic intercept and the wake function, remain unchanged. When considering boundary layers,
$\delta$ is the boundary-layer thickness. For channels, $\delta$ is the channel half-height and
we take the centreline velocity as $U_{\delta}$, to allow for direct comparison with boundary layers
\citep{Garcia-Mayoral11b,rgm2019}. Following equation \ref{eq:ud+}, the relative change in drag compared to a smooth-wall flow at the same $Re_{\tau}$ is
\begin{equation}
\frac{\Delta c_f}{c_{f0}} =  \frac{1}{(1+\Delta U^+/U_{\delta 0}^+)^2} - 1,
\label{eq:ud+2}
\end{equation}
where the subscript ‘0’ denotes reference smooth-wall values and $\Delta c_f = c_f - c_{f0}$. 

In the limit where the size of the texture elements, $L$, is vanishingly small compared to any length scales in the overlying flow, the background turbulence does not perceive the detail of each individual texture element, and only experiences the surface in an averaged sense. Such surfaces are well suited for homogenisation and can be characterised through uniform effective boundary conditions \citep{Lacis2017, bottaro2019flow}. Modelling textured surfaces by homogeneous boundary conditions is convenient and computationally cheap, since high resolution is not required near the surface to resolve the flow around the texture elements. However, homogenisation relies on a small-parameter expansion on the ratio of $L$ to the characteristic lengthscales in the flow, so it formally ceases to apply when $L$ becomes comparable to the size of the smallest eddies in the flow, which taking for the latter the diameter of quasi-streamwise vortices would be $L^+ \sim 10$. This limit would leave out most practical applications. Here, we aim to investigate the effect on turbulence of textures beyond this vanishingly small limit, i.e. once homogenisation breaks down. As a first step, in this paper, we focus on the particularly simple case of surface textures made of alternating slip/no-slip regions, which is a common model for superhydrophobic surfaces. 

In this case, the resulting homogenised boundary conditions are $v = 0$ in the wall-normal direction and $u = \ell_x (\partial u/\partial y)$ and $w = \ell_z (\partial w/\partial y)$ in the wall-parallel directions, where $\ell_x$ and $\ell_z$ are the streamwise and spanwise slip lengths \citep{Philip72}. \citet{gg20} and \citet{ibrahim2021smooth} showed that the effect of these homogenised boundary conditions is a mere offset of the origins perceived by different flow components, while turbulence remains essentially the same as over smooth walls. A streamwise slip shifts the mean velocity profile by the slip velocity, reducing the drag, while a spanwise slip allows quasi-streamwise vortices, an essential part of the near-wall turbulent cycle, to move closer to the surface \citep{Luchini96,Min04}, increasing the drag. \citet{Fairhall18} showed that for surfaces modelled using slip lengths, $\Delta U^+$ equals the offset of the virtual origin perceived by mean flow, $\ell_x^+$, and the one perceived by turbulence, $\ell_T^+$. \citet{gg20} and \cite{ibrahim2021smooth} showed that $\ell_T^+$ is governed by the interplay between the spanwise slip, $\ell_z^+$, and the transpiration. For surfaces with no transpiration, such as the cited slip/no-slip patterns, the virtual origin of turbulence only depends on the spanwise slip length, and \citet{Fairhall18} proposed 
\begin{equation}
\ell_T^+ \approx \frac{\ell_z^+}{1+ \ell_z^+/4}.
\label{eq:ellT+}
\end{equation}
from an empirical fit of the results of \citet{Busse12}.

\citet{Seo16} investigated the limits of slip-length models for slip/no-slip textures, and showed that for texture sizes $L^+ \gtrsim 10$, the instantaneous correlation between velocity and shear at the surface was lost, which would appear to set the upper limit for the applicability of a slip-length model. \citet{Fairhall18} later investigated the correlation between surface velocity and shear using a spectral approach to discriminate between the slip lengths experienced by different lengthscales in the overlying flow. They found that even scales much larger than the texture size displayed an apparent loss of correlation. \citet{Fairhall18} argued that this observed loss of correlation of the slip length was due to the intensity of the texture-coherent flow, rather than the texture size, becoming significant. We note that \citet{Jelly14} and \citet{Turk14} reported that the texture-coherent flow becomes significant compared to the background turbulence for $L^+ \gtrsim 100$, while the effect discussed by \cite{Fairhall18} sets in for $L^+ \gtrsim 10$, and is caused by the discrete slip/no-slip pattern induced by the texture being broad-band in wavelength space, scattering the texture-coherent signal across the full range of lengthscales. In a follow-up work, \citet{Fairhall19} used the amplitude-modulated flow decomposition proposed by \citet{Abderrahaman-Elena19} to filter out the texture-coherent flow, and showed that the left-over background turbulence still exhibits a linear correlation between velocity and shear, so that a slip length can be meaningfully defined.

However, when replacing the texture by the homogeneous slip lengths perceived by the overlying turbulence, results differed for $L^+ \gtrsim 25$, as the turbulent variables become no longer smooth-wall-like. \citet{Fairhall19} suggested that the breakdown of the homogeneous, slip-length model is not caused by the breakdown of the effective boundary conditions, but by a different mechanism. 
% We note that although these observations were made in the context of drag-reducing, zero-transpiration textures, we have also observed a similar behaviour for drag-increasing surfaces such as roughness \citep{Abderrahaman-Elena19} and dense canopies \citep{sharma2020turbulent}. 
% In the case where the Reynolds stress profiles are not smooth-wall-like, discrepancy in $\Delta U^+$ between the flows over textures and slips can be investigated by integrating the streamwise momentum equation \citep{Garcia-Mayoral11b}. 
They showed that the difference in $\Delta U^+$ between the flows over their textures and the flows
with equivalent homogeneous boundary conditions arose from modifications to the Reynolds stress above
rather than directly at the surface. For the simple case of slip/no-slip textures, this is particularly
clear because the zero-transpiration boundary condition ($v = 0$) yields zero Reynolds stress, $uv$,
at the surface. For the textures of \citet{Fairhall19}, the differences in $uv$ occurred at heights
$5 \lesssim y^+ \lesssim 25$ above the surface. \citet{Fairhall19} argued that the extra Reynolds
stress over textures was caused by the non-linear interactions between the texture-coherent flow and
the background turbulence, which modify the dynamics of the latter and result in degraded drag. We
note that although these observations were made in the context of drag-reducing, zero-transpiration
textures, we have also observed a similar behaviour for drag-increasing surfaces such as roughness
\citep{Abderrahaman-Elena19}, small-size dense canopies \citep{sharma2020dense} and porous substrates \citep{Hao2024}, which
suggests that this may be a common mechanism across a wide variety of textures.
\citet{Abderrahaman-Elena19} argued that a virtual-origin framework alone could only account for
roughness functions up to $-\Delta U^+ \approx 2$, spanning the early stages of the transitionally rough regime. The same conclusion was reached in \citet{Khorasani2022}
using homogenised boundary conditions.

\rgm{In this paper, we aim to identify and characterise the physical mechanism that causes the 
flow to depart from its dynamics under equivalent boundary conditions in the presence of actual,
resolved textures for $L^+ \gtrsim 25$.
In particular, we assess if the effect of the texture can be captured by replacing it by
its corresponding homogeneous boundary conditions plus, critically, the advective terms that arise from the
existence of a texture-coherent flow and its interaction with the background turbulence. Compared to Navier-Stokes, the momentum equations for the background turbulence include then additional,
`forcing' cross advective terms. We also explore preliminarily if this can be used for predictive modelling using a priori surrogates for the texture-coherent flow.}

The paper is organised as
follows. The decomposition of the flow field into background-turbulence and texture-coherent components, 
the resulting governing equations for the mean flow and background turbulence, and the cross non-linear
terms that we will introduce in our texture-less model are presented in \S \ref{sec:theory}. The 
numerical methods are outlined in \S \ref{sec:method}. The results from the model with forcing are 
presented, analysed and compared with those of texture-resolved and homogeneous-slip simulations
in \S \ref{sec:res}. Conclusions are summarised in \S \ref{sec:con}.

%%%%%%%%%%%%%%%%%%%%%%%%%%%%%%%%%%%%%%%%%%%%%%%%%%
%%%%%%%%%%%%%%%%%%%%%%%%%%%%%%%%%%%%%%%%%%%%%%%%%%
%%%%%%%%%%%%%%%%%%%%%%%%%%%%%%%%%%%%%%%%%%%%%%%%%%
%%%%%%%%%%%%%%%%%%%%%%%%%%%%%%%%%%%%%%%%%%%%%%%%%%
\section{Flow decomposition and governing equations}
\label{sec:theory}

\subsection{Amplitude-modulated triple decomposition}
\label{triple}
% Before discussing the triple decomposition with amplitude modulation, let us first introduce the conventional triple decomposition proposed by \citet{Reynolds72} which has a simpler form and can help us illustrates the idea of flow decomposition. The triple decomposition is based on the assumption that the texture lengthscales are much smaller than those over which the background overlying shear varies. In this ‘vanishingly-small’ limit, the coherent flow can be assumed to be induced around the texture elements driven by a steady, homogeneous overlying shear. This concept was already used by \citet{Luchini91}, where the riblets of a vanishingly small size are considered and they perceive the overlying turbulence as steady and homogeneous. Within the limit, the time-scales and length-scales of the overlying background-turbulent fluctuations are much larger than those of texture-coherent flow. These fluctuations are quasi-steady and quasi-homogeneous with respect to the characteristic scales of the textures, i.e. they move slowly and over long distances. 

Conventional triple decomposition is often used to obtain a texture-coherent and a texture-incoherent flow component. It decomposes the flow into a space-time-averaged mean flow, a time-averaged component which is phase-locked to the texture, and the remaining time-space fluctuations. \rgm{Taking for instance the streamwise component of the velocity, we have}
\begin{equation}
\begin{split}
u(x,y,z,t) & = U(y) + u'(x,y,z,t)\\
         & = U(y) + \widetilde{u}_u(\widetilde{x},y,\widetilde{z}) + u_T(x,y,z,t),
\label{eq:Triple_decomposition}
\end{split}
\end{equation}
where $x$, $y$ and $z$ denote the streamwise, wall-normal and spanwise directions respectively, $U(y)$ is the mean velocity profile, and $u'(x,y,z,t)$ is the full fluctuating field. The latter is further decomposed into a texture-coherent component, or dispersive flow, $\widetilde{u}_u(\widetilde{x},y,\widetilde{z})$, where $\widetilde{x}$ and $\widetilde{z}$ refer to the local coordinates within the texture period, and the remaining incoherent, background turbulent fluctuations $u_T (x,y,z,t)$. The texture-coherent component can be viewed as \rgm{being} driven by the existence of a large-scale (mean) streamwise flow, which we denote by the subscript in $\widetilde{u}_u$. The decomposition of equation \ref{eq:Triple_decomposition} has been commonly used to separate a texture-coherent contribution from the texture-incoherent, background turbulence \citep{cheng2002near,nikora2007double,coceal2007structure}. 

\begin{figure}
  \centerline{\includegraphics[width=0.8\textwidth]{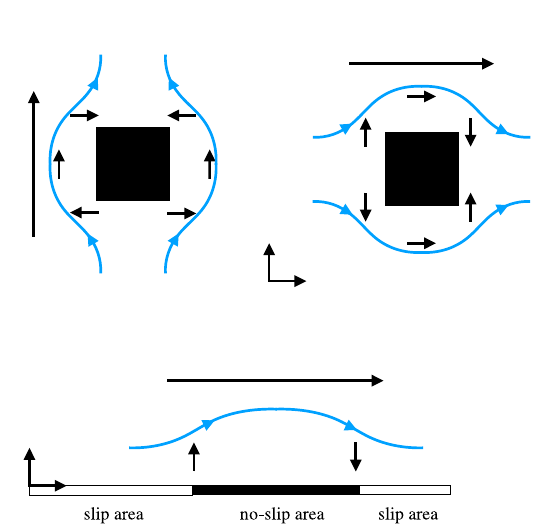}}
\mylab{15mm}{46mm}{\rotatebox{90}{\normalsize overlying streamwise velocity, $U+u_T$}}
\mylab{71mm}{96mm}{\normalsize overlying  spanwise velocity, $w_T$}
\mylab{30mm}{35mm}{\normalsize overlying streamwise (cf. spanwise) velocity}
\mylab{24.5mm}{13mm}{\large$x$ (cf. $z$)}
\mylab{12mm}{20mm}{\large$y$}
\mylab{47.5mm}{15mm}{\large$\widetilde{v}_u$ (cf. $\widetilde{v}_w$)}
\mylab{78.5mm}{15mm}{\large$\widetilde{v}_u$ (cf. $\widetilde{v}_w$)}
\mylab{19mm}{72.5mm}{\large$\widetilde{u}_u$}
\mylab{47.5mm}{72.5mm}{\large$\widetilde{u}_u$}
    \mylab{28mm}{62.5mm}{\large$\widetilde{w}_u$}
\mylab{26.5mm}{83.5mm}{\large$\widetilde{w}_u$}
\mylab{51mm}{59.5mm}{\large$x$}
\mylab{61mm}{51.5mm}{\large$z$}
\mylab{66mm}{72mm}{\large$\widetilde{u}_w$}
\mylab{90mm}{72mm}{\large$\widetilde{u}_w$}
\mylab{77mm}{53mm}{\large$\widetilde{w}_w$}
\mylab{76mm}{82.5mm}{\large$\widetilde{w}_w$}
\caption{Sketch of the texture-coherent flow induced by an overlying velocity in the streamwise
or spanwise direction. The blue lines represent streamlines of the induced flow.}
\label{fig:coherent_schematics}
\end{figure}

\citet{Abderrahaman-Elena19}, however, showed that this conventional triple decomposition does not
produce a $u_T$ free of coherence with the surface texture. They argued that the texture-coherent flow occurs in
response to the background flow, and for any given texture element it would then be driven by the full local signal of the overlying flow, which might be more or less intense than the mean $U(y)$. The texture-coherent flow would thus be modulated in intensity by the local background turbulence.
\rgm{In the above equation \ref{eq:Triple_decomposition}, the texture-coherent streamwise velocity, $\widetilde{u}_u$, driven by an
overlying streamwise velocity,} would be modulated by the streamwise background turbulence, giving
\begin{equation}
u(x,y,z,t) \approx U(y)+ u_T(x,y,z,t) + \frac{U(y) + u_T(x,y,z,t) }{U(y)}\widetilde{u}_u(\widetilde{x},y,\widetilde{z}).
\label{eq:u_decomposition}
\end{equation}
In general, \rgm{however,} the overlying flow induces a velocity field around texture elements in all three
directions, while itself has also components in all three directions.
The spanwise background flow would for instance induce a texture-coherent, spanwise
$\widetilde{w}_w$. Therefore, all three texture-coherent velocity components would have contributions
induced by all three components of the overlying flow, \rgm{such that equation \ref{eq:u_decomposition} would have additional terms for $\widetilde{u}_w$ and $\widetilde{u}_v$, the streamwise velocities induced by the background $w_T$ and $v_T$}. To illustrate this idea, figure
\ref{fig:coherent_schematics} shows a sketch of the streamwise and spanwise overlying velocities
inducing coherent flow. Applied to all velocity components, the amplitude-modulated triple
decomposition of \citet{Abderrahaman-Elena19} can be written as the product of a matrix and
a vector,
\begin{equation}
\boldsymbol{u} = \begin{bmatrix}
\widetilde{u}_u/U +1    & \widetilde{u}_v/v^*        & \widetilde{u}_w/w^* \\
\widetilde{v}_u/U       & \widetilde{v}_v/v^* + 1    & \widetilde{v}_w/w^* \\
\widetilde{w}_u/U       & \widetilde{w}_v/v^*        & \widetilde{w}_w/w^* + 1
\end{bmatrix} \begin{pmatrix}
U+u_T\\
v_T\\
w_T\\
\end{pmatrix},
\label{eq:mod_decomp}
\end{equation}
where $w^*$ and $v^*$  denote the direction-conditional averages of $w$ and $v$ over individual
roughness elements, thus giving a norm for the intensity of the respective texture-coherent
velocities. \citet{Abderrahaman-Elena19} showed that this modified form of the triple decomposition
was more effective at removing the footprint of the texture from $\boldsymbol{u_T}$. 
% $u_T$, $v_T$ and $w_T$ will be then referred to as the `turbulent' components in the following discussions.

%%%%%%%%%%%%%%%%%%%%%%%%%%%%%%%%%%%%%%%%%%%%%%%%%%%%%%%%%%%%%%%%%%%%%%%%%%%%%%%%
%%%%%%%%%%%%%%%%%%%%%%%%%%%%%%%%%%%%%%%%%%%%%%%%%%%%%%%%%%%%%%%%%%%%%%%%%%%%%%%%
%%%%%%%%%%%%%%%%%%%%%%%%%%%%%%%%%%%%%%%%%%%%%%%%%%%%%%%%%%%%%%%%%%%%%%%%%%%%%%%%
%%%%%%%%%%%%%%%%%%%%%%%%%%%%%%%%%%%%%%%%%%%%%%%%%%%%%%%%%%%%%%%%%%%%%%%%%%%%%%%%

\subsection{Governing equations for the background turbulence in flows over textures}
\label{sec:equations}
Let us now derive the governing equations for the background-turbulence component, and in particular the terms that account for the non-linear interaction with the texture-coherent flow. In a texture-resolving DNS, the incompressible flow within the channel is governed by continuity, $\nabla \cdot \boldsymbol{u} = 0$, and the Navier-Stokes equations,
\begin{equation}
% \nabla \cdot \boldsymbol{u}  = 0.
\partial_t \boldsymbol{u} +  \nabla \cdot \boldsymbol{u} \boldsymbol{u}  =  -\nabla p + \nu \nabla^2 \boldsymbol{u},
\label{conti}
\end{equation}
where $\boldsymbol{u}$ is the velocity vector with components $u$, $v$, $w$. Using the velocity decompositions introduced in section \ref{triple}, the governing momentum equations for the background turbulence can be derived by subtracting the momentum equation for texture-coherent flow from the triple-decomposed Navier-Stokes equations. We apply this first for the case of conventional triple decomposition, as a simpler example that serves us to illustrate the concept, and then apply it for the case of amplitude-modulated decomposition. 
%%%%%%%%%%%%%%%%%%%%%%%%%%%%%%%%%%%%%%%%%%%%%%%%%%%%%%%%%%%%%%%%%%%%%%%%%%%%%%%%
%%%%%%%%%%%%%%%%%%%%%%%%%%%%%%%%%%%%%%%%%%%%%%%%%%%%%%%%%%%%%%%%%%%%%%%%%%%%%%%%
\subsubsection{Texture-coherent flow obtained from conventional triple decomposition}
\label{subsec:Reynolds_decomp}

\begin{figure}
\vspace*{2mm}
  \centering
  \includegraphics[width=0.55\textwidth]{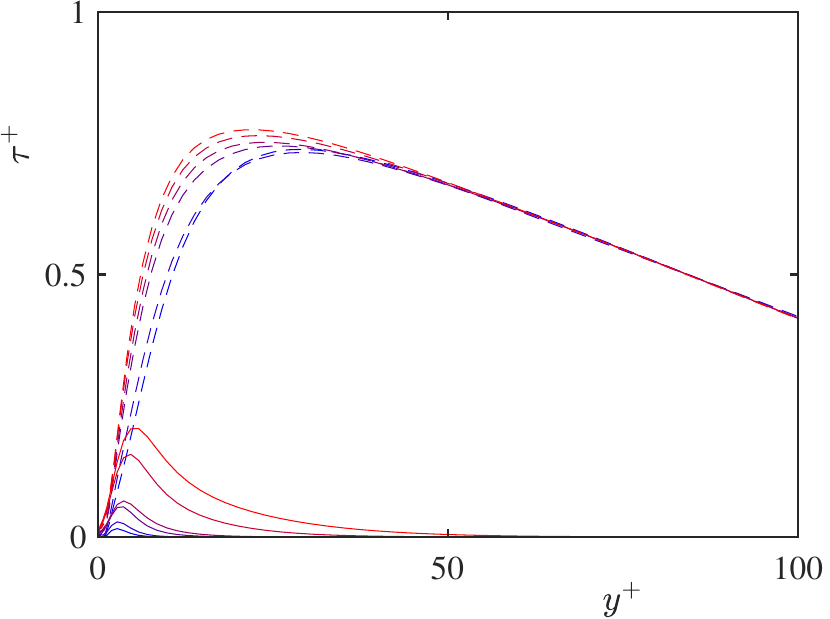}
\caption{Reynolds and dispersive stress for slip/no-slip collocated square posts at $Re_{\tau} \approx 180$. \rgm{Blue to red}, $L^+ \approx 18$, 24, 35, 47, 71 and 94.  \protect\blackdashedline, total shear Reynolds stress; \protect\blackline, dispersive stress.}
\label{fig:dispersive}
\end{figure}

Using conventional triple decomposition, the velocity vector can be written as
\begin{equation}
\boldsymbol{u} = \begin{pmatrix}
U+\widetilde{u}_u+u_T\\
\widetilde{v}_u+v_T\\
\widetilde{w}_u+w_T\\
\end{pmatrix},
\label{reynolds_v_decomp}
\end{equation}
and the pressure as $p = P + \tilde{p} + p_T$, where $P$ is the contribution that provides the mean pressure gradient driving the flow. Substituting the triple-decomposed velocities and pressure into the Navier-Stokes equations and taking the temporal and spatial average, we obtain the momentum equation for the mean flow $\boldsymbol{U} = (U, 0, 0)$,
\begin{equation}
\nabla \cdot  \left< \overline{ \boldsymbol{u_T}  \boldsymbol{u_T} }  \right> + \nabla \cdot \left< \boldsymbol{\widetilde{u}_u} \boldsymbol{\widetilde{u}_u} \right> = - \nabla P + \nu \nabla^2 \boldsymbol{U},
\label{eq:temp_spa_mean}
\end{equation}
where $\left< \left( \cdot \right) \right>$ denotes spatial averaging in the wall-parallel
directions and $\overline{\left( \cdot \right)}$  denotes temporal averaging. This has the usual
form of the mean-flow equation for rough-wall flows, with the Reynolds and the dispersive stress
on left hand side. While the dispersive stress can be significant for other textures, for the present slip/no-slip textures it is negligible compared to the Reynolds stress at least for $L^+ \lesssim 50$, only becoming relevant for texture sizes $L^+ \gtrsim 70$, as shown in figure \ref{fig:dispersive}.

The momentum equation for the texture-coherent fluctuation, $\boldsymbol{\widetilde{u}_u}$, can be obtained by taking the temporal average of the triple-decomposed Navier-Stokes equations and subtracting the momentum equation for $\boldsymbol{U}$, equation \ref{eq:temp_spa_mean}, which gives
\begin{equation}
\boldsymbol{N_c} =  -\nabla \widetilde{p} + \nu \nabla^2 \boldsymbol{\widetilde{u}_u},
\label{eq:coh_fluc}
\end{equation}
where
\begin{equation}
\boldsymbol{N_c} = \nabla \cdot  (\boldsymbol{U} + \boldsymbol{\widetilde{u}_u})  (\boldsymbol{U} + \boldsymbol{\widetilde{u}_u})  - \nabla \cdot \left< \boldsymbol{\widetilde{u}_u} \boldsymbol{\widetilde{u}_u} \right> + \nabla \cdot  \overline{ \boldsymbol{u_T}  \boldsymbol{u_T} } - \nabla \cdot  \left< \overline{ \boldsymbol{u_T}  \boldsymbol{u_T} }  \right>.
\label{N_c}
\end{equation}
Given that the background turbulence is homogeneous in the wall-parallel directions and
statistically steady in time, we have $\nabla \cdot  \overline{ \boldsymbol{u_T}
\boldsymbol{u_T} } - \nabla \cdot  \left< \overline{ \boldsymbol{u_T}  \boldsymbol{u_T} }
\right> \approx 0$ in the non-linear term $\boldsymbol{N_c}$, and the resulting momentum
equation for the texture-coherent fluctuation, $\boldsymbol{\widetilde{u}_u}$, simplifies then to
\begin{equation}
\nabla \cdot  (\boldsymbol{U} + \boldsymbol{\widetilde{u}_u})  (\boldsymbol{U} + \boldsymbol{\widetilde{u}_u})  - \nabla \cdot \left< \boldsymbol{\widetilde{u}_u} \boldsymbol{\widetilde{u}_u} \right> =  -\nabla \widetilde{p} + \nu \nabla^2 \boldsymbol{\widetilde{u}_u}.
\label{eq:coh_fluc_1}
\end{equation}

Finally, the momentum equation for the background turbulence can be obtained by subtracting
the above momentum equation for $\boldsymbol{\widetilde{u}_u}$ from the full triple-decomposed
Navier-stokes equation, yielding
\begin{equation}
\partial_t \boldsymbol{u_b} + \nabla \cdot \boldsymbol{u_b} \boldsymbol{u_b} +  \boldsymbol{N_b}  = -\nabla p_b + \nu \nabla^2 \boldsymbol{u_b},
\label{eq:background}
\end{equation}
where $\boldsymbol{u_b} = \boldsymbol{U} + \boldsymbol{u_T}$ includes the mean $\boldsymbol{U}$ profile and the background turbulent fluctuations, and is thus the flow we are interested in solving for in our texture-less DNSs, and ${p_b}$ is the corresponding background pressure, ${p_b} = P + p_T$. The additional non-linear terms, $\boldsymbol{N_b}$, are
\begin{equation}
\begin{split}
\boldsymbol{N_b} 
=  \nabla \cdot  \boldsymbol{\widetilde{u}_u}  \boldsymbol{u_T} + \nabla \cdot \boldsymbol{u_T}  \boldsymbol{\widetilde{u}_u} + \nabla \cdot \left< \boldsymbol{\widetilde{u}_u} \boldsymbol{\widetilde{u}_u} \right>,
\end{split}
\label{eq:n_t}
\end{equation}
where $\nabla \cdot \boldsymbol{\widetilde{u}_u}  \boldsymbol{u_T} + \nabla \cdot \boldsymbol{u_T}  \boldsymbol{\widetilde{u}_u} $ are the non-linear interactions of the background turbulence and the texture-coherent flow and $\nabla \cdot \left< \boldsymbol{\widetilde{u}_u} \boldsymbol{\widetilde{u}_u} \right>$ is the dispersive stress, which acts only on the mean flow $\boldsymbol{U}$. The presence of surface texture thus modifies the background turbulence through these non-linear terms. 

%%%%%%%%%%%%%%%%%%%%%%%%%%%%%%%%%%%%%%%%%%%%%%%%%%%%%%%%%%%%%%%%%%%%%%%%%%%%%%%%
%%%%%%%%%%%%%%%%%%%%%%%%%%%%%%%%%%%%%%%%%%%%%%%%%%%%%%%%%%%%%%%%%%%%%%%%%%%%%%%%
\subsubsection{Texture-coherent flow obtained from amplitude-modulated triple decomposition}
\label{subsec:mod_decomp}

Let us now derive the momentum equations for the background turbulence using the
amplitude-modulated triple decomposition of equation \ref{eq:mod_decomp}. For the present
case of slip/no-slip textures, we make the following simplifications. At the surface, the
wall-normal velocity $v$ is zero throughout, and therefore any flow induced by the
background $v$ is small, as reported in \citet{Fairhall19}, so we neglect $\widetilde{u}_v$,
$\widetilde{v}_v$ and $\widetilde{w}_v$ in equation \ref{eq:mod_decomp}. Let us however
note that this $v$-induced flow could be important for protruding textures such as pyramids,
cones, and roughness in general. In addition, we also neglect the flow induced by the
background $w$, since the latter is weak compared to the background streamwise flow for
slip/no-slip textures, as also reported in \citet{Fairhall19}. The dominant texture-coherent
effect is then the flow induced by streamwise velocity $U+u_T$, and the modulated triple
decomposition simplifies then to 
\begin{equation}
\begin{split}
\boldsymbol{u} &= \left ( \boldsymbol{I} + \begin{bmatrix}
\widetilde{u}_u/U     & 0       & 0\\
\widetilde{v}_u/U       & 0       & 0\\
\widetilde{w}_u/U       & 0       & 0
\end{bmatrix} \right ) \cdot \begin{pmatrix}
U+u_T\\
v_T\\
w_T\\
\end{pmatrix},\\
&= \left ( \boldsymbol{I} + \boldsymbol{C} \right ) \cdot \boldsymbol{u_b},
\end{split}
\label{V2}
\end{equation}
where we assume that $\widetilde{\boldsymbol{u}}_u$ can still be obtained by ensemble
averaging \citep{Abderrahaman-Elena19, Fairhall19} and is therefore still governed by
equation \ref{eq:coh_fluc}. Similarly, the pressure is decomposed into 
\begin{equation}
p = P + p_T + \frac{U+u_T}{U} \widetilde{p}_u, 
\label{V2-p}
\end{equation}
which can be further written as
\begin{equation}
p = p_b + \boldsymbol{Q} \cdot \boldsymbol{u_b}, 
\label{V2-p2}
\end{equation}
where $\boldsymbol{Q} = [\widetilde{p}_u/U \quad 0 \quad 0]$, corresponding to a linear expansion about $U$.
% \hl{(we only do the modulation to the non-linear term)}

The Navier-Stokes equations can then be written as
\begin{equation}
\begin{split}
\partial_t \left( \left( \boldsymbol{I} + \boldsymbol{C} \right) \cdot \boldsymbol{u_b} \right) + & \nabla \cdot  \left[ \left( \left( \boldsymbol{I}+\boldsymbol{C} \right) \cdot \boldsymbol{u_b} \right)  \left( \left( \boldsymbol{I}+\boldsymbol{C} \right) \cdot \boldsymbol{u_b} \right) \right] \\
& = -\nabla \left( p_b + \boldsymbol{Q} \cdot \boldsymbol{u_b} \right)  + \nu \nabla^2 \left( \left( \boldsymbol{I}+ \boldsymbol{C} \right) \cdot \boldsymbol{u_b} \right),
\label{eq:all}
\end{split}
\end{equation}

Following the procedure in section \ref{subsec:Reynolds_decomp}, the momentum equation for the background flow can be obtained by subtracting the governing equation for texture-coherent flow, equation \ref{eq:coh_fluc}, from equation \ref{eq:all}, 
\begin{equation}
\partial_t \boldsymbol{u_b} + \nabla \cdot  \boldsymbol{u_b} \boldsymbol{u_b} + \boldsymbol{N_b'}  = -\nabla p_b + \nu \nabla^2 \boldsymbol{u_b} + \boldsymbol{R},
\label{eq:bgd_2}
\end{equation}
where 
\begin{equation}
\begin{split}
\boldsymbol{N_b'} 
% &= \partial_t \left( \boldsymbol{C} \cdot \boldsymbol{u_b} \right) \\
&= \nabla \cdot \left[ \left(\boldsymbol{C} \cdot \boldsymbol{u_b} \right) \boldsymbol{u_b} + \boldsymbol{u_b} \left(\boldsymbol{C} \cdot \boldsymbol{u_b} \right)\right] + \nabla \cdot \left[ \left( \boldsymbol{C} \cdot \boldsymbol{u_b} \right) \left( \boldsymbol{C} \cdot \boldsymbol{u_b} \right) \right] \\
& - \nabla \cdot  (\boldsymbol{U} + \boldsymbol{\widetilde{u}_u})  (\boldsymbol{U} + \boldsymbol{\widetilde{u}_u})  + \nabla \cdot \left< \boldsymbol{\widetilde{u}_u} \boldsymbol{\widetilde{u}_u} \right>,
% &+ \nabla \left( \boldsymbol{Q} \cdot \boldsymbol{u_b} \right) - \nabla \widetilde{p} \\
% &- \nu \nabla^2 \left( \boldsymbol{C} \cdot \boldsymbol{u_b} \right) + \nu \nabla^2 \boldsymbol{\widetilde{u}_u} .
\end{split}
\label{eq:nb'}
\end{equation}
and
\begin{equation}
\begin{split}
\boldsymbol{R} 
= \partial_t \left( \boldsymbol{C} \cdot \boldsymbol{u_b} \right)
% &+ \nabla \cdot \left[ \left(\boldsymbol{C} \cdot \boldsymbol{u_b} \right) \boldsymbol{u_b}^T + \boldsymbol{u_b} \left(\boldsymbol{C} \cdot \boldsymbol{u_b} \right)^T\right] + \nabla \cdot \left[ \left( \boldsymbol{C} \cdot \boldsymbol{u_b} \right) \left( \boldsymbol{C} \cdot \boldsymbol{u_b} \right)^T \right] \\
% & - \left (\nabla \cdot  (\boldsymbol{U} + \boldsymbol{\widetilde{u}_u})  (\boldsymbol{U} + \boldsymbol{\widetilde{u}_u})^T  - \nabla \cdot \left< \boldsymbol{\widetilde{u}_u} \boldsymbol{\widetilde{u}_u}^T \right> \right )\\
+ \nabla \left( \boldsymbol{Q} \cdot \boldsymbol{u_b} \right) - \nabla \widetilde{p}
- \nu \nabla^2 \left( \boldsymbol{C} \cdot \boldsymbol{u_b} \right) + \nu \nabla^2 \boldsymbol{\widetilde{u}_u} .
\end{split}
\label{eq:r}
\end{equation}

Here, we have drawn a parallel with the result using conventional triple decomposition of equation \ref{eq:background} by grouping the additional advective terms into $\boldsymbol{N_b'}$, and any other additional terms arising from the amplitude modulation into a residual $\boldsymbol{R}$. In appendix \ref{appResid} we report on the magnitude of the latter, and show that $\boldsymbol{N_b'}$ is the dominant term. We will thus neglect the residual $\boldsymbol{R}$ in our model.

\begin{figure}
  \centering
  \includegraphics[trim=0 0 0 0,clip,width=\textwidth]{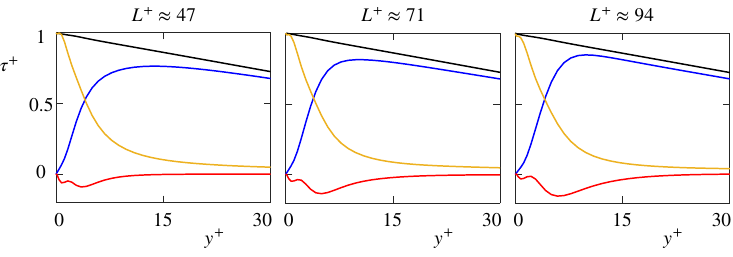}
\caption{ Comparison of viscous stress (\protect\yellowline), shear Reynolds stress (\protect\blueline), averaged forcing term (\protect\redline) and total stress (\protect\blackline) for FA47, FA71 and FA94.}
\label{fig:tot_stress}
\end{figure}

Finally, we can derive the momentum equation for the mean velocity $\boldsymbol{U}$ analogously to equation
\ref{eq:temp_spa_mean} by taking the temporal average of  equation \ref{eq:bgd_2} particularised for the
\rgm{Fourier $x$-$z$ zero mode, that is, for the $x$-$z$ spatial average}. This gives
% the simulations with relatively large texture sizes $L^+ \approx$ 47, 71 and 94, the momentum equation for the mean velocity $U$ can be written as
\begin{equation}
\nabla \cdot \left< \overline{\boldsymbol{u_T}  \boldsymbol{u_T}} \right> + \boldsymbol{N_m}= - \nabla P + \nu \nabla^2 \boldsymbol{U},
\label{eq:temp_spa_mean_3}
\end{equation}
where the forcing term $\boldsymbol{N_m}$ is
\begin{equation}
\boldsymbol{N_m} = \nabla \cdot \left< \boldsymbol{\widetilde{u}_u} \boldsymbol{\widetilde{u}_u} \right> +
\nabla \cdot \left< \boldsymbol{C}  \cdot \overline{\boldsymbol{u_T}  \boldsymbol{u_T}}  + \overline{\boldsymbol{u_T}  \boldsymbol{u_T}}  \cdot \boldsymbol{C}^T \right> + \nabla \cdot \left< \boldsymbol{C}  \cdot  \overline{\boldsymbol{u_T}  \boldsymbol{u_T}}  \cdot \boldsymbol{C}^T \right>.
\label{eq:temp_spa_mean_F}
\end{equation}
Here, $\boldsymbol{u_T} \left(\boldsymbol{C} \cdot\boldsymbol{u_T}\right)$ has been rearranged as
$\boldsymbol{u_T} \boldsymbol{u_T} \cdot\boldsymbol{C}^T$ before averaging in time.
Integrating the above mean momentum equation in $y$ gives the following stress balance
\begin{equation}
\nu \frac{d U}{d y} - \left< \overline{u_T v_T} \right> -  N_{m,x} = u_{\tau}^2 \frac{\delta'- y}{\delta'},
\label{eq:mean_mome_forcing}
\end{equation}
where the role of the dispersive Reynolds stress in equation \ref{eq:temp_spa_mean} is now played by an extended term $N_{m,x}$,
\begin{equation}
N_{m,x} = \left< \widetilde{u}_u \widetilde{v}_u + \frac{\widetilde{u}_u \widetilde{v}_u }{U^2}  \overline{u_T v_T}  + \frac{\widetilde{u}_u }{U} \overline{u_T v_T}  + \frac{\widetilde{v}_u }{U} \overline{u_T u_T}  \right>.
\label{eq:mean_mome_forcing_N}
\end{equation}
$N_{m,x}$ includes contributions from the background turbulence, and can thus only be calculated from the 
results in \S \ref{sec:res} a posteriori in order to verify that the sum of the three stresses in equation
\ref{eq:mean_mome_forcing} is indeed linear. Results
of such measurement are portrayed for reference in 
figure \ref{fig:tot_stress} for the collocated textures of sizes $L^+\gtrsim50$, as for smaller textures the
contribution of $N_{m,x}$ is negligible. For the sizes
portrayed its intensity increases with $L^+$, although it remains small compared to the viscous and background
Reynolds stress.

%%%%%%%%%%%%%%%%%%%%%%%%%%%%%%%%%%%%%%%%%%%%%%%%%%%%%%%%%%%%%%%%%%%%%%%%%%%%%%%%
%%%%%%%%%%%%%%%%%%%%%%%%%%%%%%%%%%%%%%%%%%%%%%%%%%%%%%%%%%%%%%%%%%%%%%%%%%%%%%%%
%%%%%%%%%%%%%%%%%%%%%%%%%%%%%%%%%%%%%%%%%%%%%%%%%%%%%%%%%%%%%%%%%%%%%%%%%%%%%%%%
%%%%%%%%%%%%%%%%%%%%%%%%%%%%%%%%%%%%%%%%%%%%%%%%%%%%%%%%%%%%%%%%%%%%%%%%%%%%%%%%
\section{Numerical Method}
\label{sec:method}
% 1. common features - from Fairhall 19
% 2. particular features for simulations with forcing terms
% 3. resolution etc
% 4. list of simulations, specifying which are from previous works. 

% go general to particular. short describing the general features of the code/setup that apply to all simulaions. then detail what is specific for texture resolved simulations, then what'sn specific for slip sims., then for forcing simulations. 

% then give particular parameters of the simulations why you have chosen to run those, and specify which simulations are new and which which from previous work.

To investigate the non-linear interaction between background turbulence and texture-coherent flow,
direct numerical simulations (DNS) of turbulent channels were conducted. The numerical code
is adapted from that of \citet{Fairhall19}, \rgm{and is} briefly summarised here. The three-dimensional incompressible Navier-Stokes equations are solved using a spectral discretisation in the
streamwise and spanwise directions with the wall-normal direction discretised by second-order
finite differences on a staggered grid. A fractional step method \citep{Kim85}, combined with
a three-step Runge-Kutta scheme, is used to advance in time, with a semi-implicit scheme used
for the viscous terms and an explicit scheme for the advective terms \citep{Le91}. The
simulations were run with constant mean pressure gradients, adjusted to achieve the desired
friction Reynolds numbers, Re$_{\tau}$. The channel is of size $2\pi \delta \times \pi \delta
\times 2 \delta$ in the streamwise, spanwise and wall-normal directions respectively, where
$\delta$ is the channel half-height. \rgm{Once a statistically steady state was reached, statistics were obtained over 10 times the
characteristic largest-eddy-turnover time, $\delta/u_\tau$.}

% which are texture-resolving simulations, homogeneous slip-length simulations as well as homogeneous slip-length simulations plus forcing terms introduced in the previous section. 
Three sets of DNSs were conducted. The first set consists of texture-resolving simulations
with patterns of alternating regions of slip/no-slip boundary conditions on the channel walls.
The first set used the DNS code from \citet{Fairhall19} unmodified. The surfaces are assumed
rigidly flat, which leads to zero transpiration. This is a widely used idealisation for
superhydrophobic surfaces, where the free-slip regions represent the gas pockets, and the
no-slip regions the exposed tips of solid posts. The assumption of free slip is reasonable
if the gas pockets are sufficiently deep \citep{schonecker2014influence}. The assumption of
flat, rigid interfaces is reasonable for $L^+ \lesssim 30$ for typical applications
\citep{seo2018effect}. We note nevertheless that in this study we use this model to keep the
surface texture as simple as possible, regardless of whether the model is a suitable
representation or not of superhydrophobic textures. The latter question is out of the scope
of this work.

\begin{figure}
  \centering
  \includegraphics[trim=0 0 0 0,clip,width=0.9\textwidth]{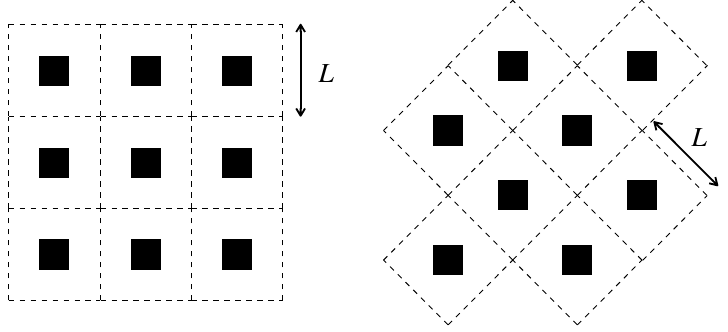}
\caption{Schematic of collocated- and staggered-post arrangements for periodic textures.}
\label{fig:text_arr}
\end{figure}

We consider texture elements consisting of periodic arrays of square posts, both in
collocated and staggered arrangements, as illustrated in figure \ref{fig:text_arr}, with
a solid fraction, the ratio of post area to total surface area, $\phi_s \approx 1/9$. The
texture unit is a square of side $L$, repeated periodically in the streamwise and spanwise
directions. For our different simulations, the texture size ranges from $L^+ \approx$ 18
to 94 for the collocated textures and $L^+ \approx$ 33 to 66 for the staggered textures.
% where the data for the texture-resolving simulations for collocated textures $L^+ \approx$ 18 to 47 are from \citet{Fairhall19}. We adapted their code and conducted additional simulations with larger texture sizes $L^+ \approx $ 71 and 94 to extend the database. The solver for texture-resolving simulations were validated by \citet{Fairhall19}.  
% For staggered arrangement, the texture has approximately the same solid fraction as the collocated pattern, and the texture size ranges from  $L^+ \approx$ 33 to 66. 

The second set of DNSs replaces the textured surfaces by their corresponding homogeneous
slip lengths, obtained a posteriori from the corresponding texture-resolving simulations. 
This set also used the DNS code from \citet{Fairhall19} unmodified. The last set of DNSs are
homogeneous-slip simulations with added forcing terms, with the same slip boundary conditions
of the second set. The governing momentum equations have the additional non-linear forcing terms introduced in section \ref{sec:equations}, with the texture-coherent velocities obtained
from ensemble-averaging in the texture-resolving simulations. 

In order to reduce the computational cost of the texture-resolving simulations, the code uses
a multiblock layout, where the near-wall regions have finer grids compared to the channel
centre. This is done in order to resolve fully not just the channel turbulence but also
the texture-coherent flow, which typically requires higher resolution. In the blocks containing the walls, each texture element is resolved by $24 \times 24$ grid points in the $x$ and $z$ directions. In the central block, the grid resolution is $\Delta x^+ \approx$ 8.8 and
$\Delta z^+  \approx$ 4.4. The latter resolution is also used throughout in the simulations with homogeneous slip. In the wall-normal direction, the grid is stretched, with resolution $\Delta y_{min}^+ \approx$ 0.3 at the surfaces and $\Delta y_{max}^+ \approx$ 3 in the channel centre,
\rgm{and with the fine-$x$-$z$- resolution blocks extending a height $\approx L$ into the channel from the wall, which was verified a posteriori to exceed the height at which any fine, texture-induced flow became vanishingly small.
Overall, this resulted for instance in roughly $10^8$ grid points for case TX18 or $2\times10^7$ for TX47, which can be compared with $2\times10^6$ for the 
corresponding smooth-wall and texture-less simulations, or $4\times10^8$ for TX18 if uniform resolution in $x$ and $z$ had been applied throughout.}
% Full details of the simulation parameters are also given in table \ref{tab:simulations}. 

% In this section, we numerically solve the governing equations (\ref{eq:fluc_2}-\ref{equ:Nu_2}) derived in section \ref{sec:equations} 
% The homogeneous slip-length simulations plus forcing terms were conducted with a similar setups, where slip-length boundary conditions are applied on the top and bottom wall of a turbulent channel. The governing momentum equations of this set of simulations have additional non-linear forcing terms as introduced in section \ref{sec:equations}. Same to the slip lengths, the components of texture-coherent flow velocity used to form the forcing terms are also obtained \textit{a posteriori} from texture-resolving simulations. 

The texture-resolving and homogeneous-slip simulations with collocated elements with 
sizes $L^+ \approx 18$ to 47 are from \citet{Fairhall19}. Additional simulations with
$L^+ \approx 71$ and 94, as well as the simulations with staggered elements, have been
conducted to complete and expand the database. Simulations with homogeneous slip and added
forcing have been conducted matching all the above cases to investigate the effect of
the non-linear interaction. The parameters of all three sets of simulations are listed
in table \ref{tab:simulations}.

\begin{table}
 \begin{center}
\vspace*{-2mm}
  \begin{tabular}{lrrrrrrrrrr}
  %{L{12mm}R{9mm}R{9mm}R{9mm}R{9mm}R{9mm}R{9mm}R{9mm}R{9mm}R{9mm}R{9mm}}
      Case\hspace{6mm} & \hspace{5mm}$Re_{\tau}$\hspace{1mm} & \hspace{5mm}$L^+$ & \hspace{4mm}$N_{x,w}$ & \hspace{4mm}$N_{z,w}$\hspace{-1mm} & \hspace{7mm}$\ell_x^+$ & \hspace{7mm}$\ell_z^+$ & \hspace{5mm}$\Delta U^+\!\!\!\!\!$\\[6pt]
      TX18   & 179.7 & 17.7 & 1576 & 768 &  5.8 &  4.0 &  3.7 \\
      TX24   & 180.2 & 23.6 & 1152 & 576 &  6.9 &  4.3 &  4.1 \\
      TX35   & 179.9 & 35.3 &  768 & 384 &  8.5 &  5.0 &  4.8 \\
      TX35H  & 407.0 & 35.3 & 1728 & 864 &  8.4 &  5.1 &  5.4 \\
      TX47   & 180.1 & 47.1 &  576 & 288 & 10.0 &  6.3 &  5.6 \\
      TX71   & 180.0 & 70.7 &  384 & 192 & 12.7 &  7.3 &  7.5 \\
      TX94   & 179.7 & 94.2 &  288 & 144 & 14.6 &  8.5 &  9.1 \\
      TX94H  & 360.7 & 94.2 &  576 & 288 & 14.2 & 10.0 &  9.7 \\[2pt]

      sTX35  & 181.0 & 33.3 &  576 & 288 &  6.0 &  4.4 &  3.7 \\
      sTX47  & 180.7 & 49.9 &  384 & 192 &  7.3 &  4.7 &  4.1 \\
      sTX71  & 179.8 & 66.6 &  288 & 144 &  8.5 &  3.2 &  4.7 \\[-1pt]
      \hline\\[-8pt]
      
      SL18   & 179.9 & 17.7 &  128 & 128 &  5.8 &  4.0 &  4.0 \\
      SL24   & 179.9 & 23.6 &  128 & 128 &  6.9 &  4.3 &  5.0 \\
      SL35   & 180.3 & 35.3 &  128 & 128 &  8.5 &  5.0 &  6.5 \\
      SL35H  & 405.4 & 35.3 &  256 & 256 &  8.4 &  5.1 &  6.5 \\
      SL47   & 179.7 & 47.1 &  128 & 128 & 10.0 &  6.3 &  7.8 \\
      SL71   & 179.9 & 70.7 &  128 & 128 & 12.7 &  7.3 & 10.2 \\
      SL94   & 179.9 & 94.2 &  128 & 128 & 14.6 &  8.5 & 12.1 \\
      SL94H  & 359.7 & 94.2 &  256 & 256 & 14.2 & 10.0 & 12.1 \\[2pt]

      sSL35  & 180.1 & 33.3 &  128 & 128 &  6.0 &  4.4 &  4.2 \\
      sSL47  & 180.4 & 49.9 &  128 & 128 &  7.3 &  4.7 &  5.3 \\
      sSL71  & 180.0 & 66.6 &  128 & 128 &  8.5 &  3.2 &  6.9 \\[-1pt]
      \hline\\[-6pt]
      
      FA18   & 179.9 & 17.7 &  128 & 128 &  5.8 &  4.0 &  3.6 \\
      FA24R  & 180.1 & 23.6 &  128 & 128 &  6.9 &  4.3 &  4.8 \\
      FA24   & 180.0 & 23.6 &  128 & 128 &  6.9 &  4.3 &  4.0 \\
      FA24S  & 180.2 & 23.6 &  128 & 128 &  6.9 &  4.3 &  4.0 \\
      FA35R  & 180.1 & 35.3 &  128 & 128 &  8.5 &  5.0 &  6.0 \\
      FA35   & 179.8 & 35.3 &  128 & 128 &  8.5 &  5.0 &  4.8 \\
      FA35S  & 180.0 & 35.3 &  128 & 128 &  8.5 &  5.0 &  5.0 \\
      FA35H  & 404.7 & 35.3 &  256 & 256 &  8.4 &  5.1 &  5.4 \\
      FA47R  & 180.3 & 47.1 &  128 & 128 & 10.0 &  6.3 &  7.0 \\
      FA47   & 180.2 & 47.1 &  128 & 128 & 10.0 &  6.3 &  5.7 \\
      FA71R  & 180.0 & 70.7 &  128 & 128 & 12.7 &  7.3 &  8.9 \\
      FA71   & 180.1 & 70.7 &  128 & 128 & 12.7 &  7.3 &  7.9 \\
      FA94R  & 179.9 & 94.2 &  128 & 128 & 14.6 &  8.5 & 10.5 \\
      FA94   & 180.0 & 94.2 &  128 & 128 & 14.6 &  8.5 &  9.9 \\
      FA94H  & 359.5 & 94.2 &  256 & 256 & 14.2 & 10.0 & 10.2 \\[2pt]

      sFA35  & 179.8 & 33.3 &  128 & 128 &  6.0 &  4.4 &  3.6 \\
      sFA47  & 180.3 & 49.9 &  128 & 128 &  7.3 &  4.7 &  4.1 \\
      sFA71  & 180.1 & 66.6 &  128 & 128 &  8.5 &  3.2 &  5.4 \\[2pt]
    
  \end{tabular}
  %\vspace*{-1mm}
  \caption{Simulation parameters for the texture-resolving, slip-only and slip-plus-forcing simulations. For the case names, TX and sTX
  indicate resolved collocated and staggered textures, SL and sSL slip-only simulations, and FA and sFA simulations with forcing. The number in the case name is approximately the texture size in wall units, $L^+$, listed also for texture-less simulations as a reference to the corresponding textured ones. An appended R indicates forcing based on conventional triple decomposition; otherwise the forcing is based on amplitude-modulated decomposition. 
  An appended S indicates the use of an a priori surrogate model for the
  texture-coherent flow.
  An appended H indicates higher friction Reynolds number, $Re_{\tau}$. $N_{x,w}$ and $N_{z,w}$ are the number of grid points in the streamwise and spanwise directions in the refined blocks that contain the channel walls. $\ell_x^+$ and $\ell_z^+$ are the streamwise and spanwise slip lengths, and $\Delta U^+$ is the resulting velocity increment away from the wall.}
  \label{tab:simulations}
 \end{center}
\end{table}

\subsection{Dealiasing for the forcing terms}
When calculating the product of two variables, such as in the advective term, in discrete
Fourier space, the problem of aliasing arises. The convolution of two discrete functions,
$\hat{f}_M$ and $\hat{f}_N$, containing $M$ and $N$ discrete modes, results in a function
containing $M+N$ modes. However, if the product function allocated for the convolution,
$\hat{f}_P$, is of size $P < M+N$, the excess modes cannot be correctly represented. This
additional, high frequency information is then reflected into the resolved modes of the
product function. %This contamination of the solution is known as aliasing and needs to be
%considered for the non-linear terms.

\begin{figure}
  \centering
  \includegraphics[trim=0 0 0 0,clip,width=\textwidth]{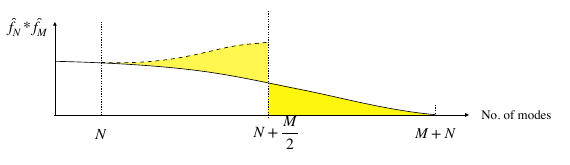}
\caption{Sketch of the dealiasing strategy for the forcing terms. The spectral region shaded in yellow with modes larger than $N+M/2$ is reflected into the shaded region under the dashed line, which includes only modes larger than $N$, and thus avoids aliasing of modes below $N$.}
\label{fig:dealiasing}
\end{figure}

The method typically used to address this aliasing is to pad $\hat{f}_M$ and $\hat{f}_N$
with additional modes, with zero value, before the multiplication \citep{canuto2012spectral}.
For the non-linear term  $ \left ( \boldsymbol{u_T} \boldsymbol{u_T} \right)$ in equation \ref{eq:n_t}, we use the standard `2/3 rule' for dealiasing. All functions have the same size,
$M=N=P$, so they need to be padded with an additional $N/2$ modes to prevent aliasing in
$\hat{f}_P$ in modes 0 to $N$. Aliasing still occurs from the reflection of modes greater than $3N/2$ into modes from $N$ to $3N/2$, but this is of no consequence as these modes are
\rgm{discarded} once the convolution product has been calculated. In contrast with the standard advection, for the product
$\boldsymbol{\widetilde{u}_u} \boldsymbol{u_T}$, the convolution of the background
turbulence and the texture-coherent flow, the sizes of the two components need not be the same.
We have for instance $N = 128$ for $\boldsymbol{u_T}$, with $M$ different for cases with
different texture size, and typically $M \geq N$. Since we are interested in solving only the background turbulence, that is, modes up to $N$, for dealiasing in this case we need at least
$P = N + M/2$, as illustrated in figure \ref{fig:dealiasing}. Taking case FA35 as an example, in the streamwise direction the texture-coherent component $\boldsymbol{\widetilde{u}_u}$ has $M = 768$ points in discrete Fourier space, so $X = 128 + 768/2 = 512$ is required for dealiasing.

%%%%%%%%%%%%%%%%%%%%%%%%%%%%%%%%%%%%%%%%%%%%%%%%%%%%%%%%%%%%%%%%%%%%%%%%%%%%%%%%
%%%%%%%%%%%%%%%%%%%%%%%%%%%%%%%%%%%%%%%%%%%%%%%%%%%%%%%%%%%%%%%%%%%%%%%%%%%%%%%%
%%%%%%%%%%%%%%%%%%%%%%%%%%%%%%%%%%%%%%%%%%%%%%%%%%%%%%%%%%%%%%%%%%%%%%%%%%%%%%%%
%%%%%%%%%%%%%%%%%%%%%%%%%%%%%%%%%%%%%%%%%%%%%%%%%%%%%%%%%%%%%%%%%%%%%%%%%%%%%%%%

\section{Results and discussion}
\label{sec:res}

\begin{figure}
  \centering
  \vspace*{2mm}
  \includegraphics[trim=0 0 0 0,clip,width=0.64\textwidth]{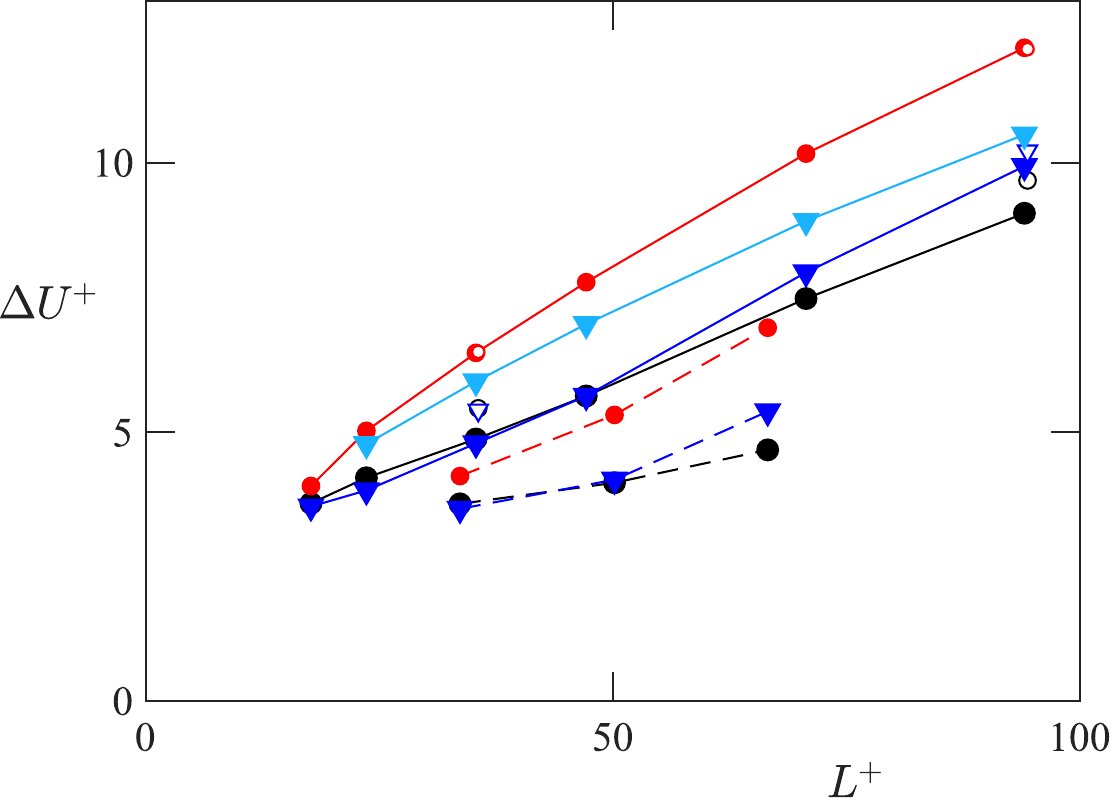}
  \hspace*{7mm}
  \vspace*{1mm}
  \caption{Comparison of $\Delta U^+$ obtained from texture-resolved, slip-only and slip-plus-forcing
simulations. Full symbols are for $Re_{\tau} \approx 180$, and open symbols for $Re_{\tau} \approx
350$--$400$. Solid and dashed lines are for simulations with collocated and staggered texture
arrangements, respectively. {\large\color{Black}$\bullet$}, texture-resolved simulations;
{\large\color{Colour2}$\bullet$}, simulations with homogeneous slip boundary conditions only;
{\large \color{Colour4}$\blacktriangledown$}, simulations with homogeneous slip plus forcing based
on conventional triple decomposition; {\large\color{Colour1}$\blacktriangledown$}, simulations with
homogeneous slip plus forcing based on amplitude-modulated triple decomposition.}
\label{fig:l_vs_delta_u}
\end{figure}

In this section, we present and discuss the results of the DNSs summarised
in table \ref{tab:simulations}. We first compare the drag predictions
obtained from simulations with the texture resolved and those with
its effect modelled. Figure \ref{fig:l_vs_delta_u} portrays those results
in terms of the velocity increment $\Delta U ^+$. For each simulation
setup, the figure shows the usual increase of $\Delta U^+$ with texture
size $L^+$. The results with resolved textures and with the corresponding
slip boundary conditions agree well only up to texture sizes $L^+ \approx 20$ 
\citep[for smaller textures see results in][]{Fairhall19}
for collocated layouts and $L^+\approx35$ for staggered ones, although we note that
the streamwise spacing between successive rows of posts is then $L_x^+ \approx 25$,
close to the collocated value. For larger spacings, using the equivalent
homogeneous slip increasingly overpredicts $\Delta U ^+$, by $\sim35$\% for $L^+\gtrsim35$
in the case of collocated posts, and by a similar proportion for staggered ones.
This difference was already reported by \cite{Fairhall19}. They argued that,
given that the effective boundary conditions perceived by the overlying turbulence were the same for
texture-resolved and slip simulations, the difference in drag
had to arise from differences in the overlying flow. Introducing the forcing terms
without amplitude modulation from \S \ref{subsec:Reynolds_decomp} improves the prediction
of $\Delta U ^+$ for slip-only simulations partially, as shown in Figure \ref{fig:l_vs_delta_u}.
The error remains however large, and is only reduced by 30--50\%.
In turn, introducing the forcing terms with amplitude modulation from \S \ref{subsec:mod_decomp}
yields values of $\Delta U ^+$ in good agreement with those of texture-resolved simulations at least up to
texture sizes $L^+\approx70$. As the texture size increases further to $L^+\approx100$, the results begin to depart.
Although the deviation remains below 10\%, we already expected the forcing model to break down in this
range of $L^+$, as the assumption gradually ceases to hold that there is sufficient separation of scales between the
overlying shear and the texture-coherent flow it induces, rendering the flow decomposition used in the forcing model invalid. This is further discussed in \S \ref{sec:limit}.

For the staggered-posts layouts, the values of $\Delta U ^+$ are lower than for collocated layouts
with the same $L^+$. This is in agreement with the Stokes-flow predictions for small textures of
\cite{sbragaglia2007effective}. They argued that collocated textures channel the flow through 
streamwise-aligned channels between posts, while staggered arrangements obstruct this channelling
effect, reducing the mean slip velocity. This obstruction was also important in the simulations of \cite{seo2018effect}, 
who observed that the slip lengths measured from DNSs of surfaces with randomly distributed texture
elements was reduced by approximately 30\% compared to collocated elements. In any event, the
results for texture-resolved, slip-only and slip-plus-forcing simulations for staggered posts
follow generally the same trends observed for collocated posts, albeit with lower values of $\Delta U ^+$ 
for the same $L^+$, and are therefore not presented in the remainder of this section. They are
nevertheless included for completeness in appendix \ref{appStagg}. 

\begin{figure}
  \centering
   \vspace*{2mm}
 \includegraphics[trim=0 0 0 0,clip,width=\textwidth]{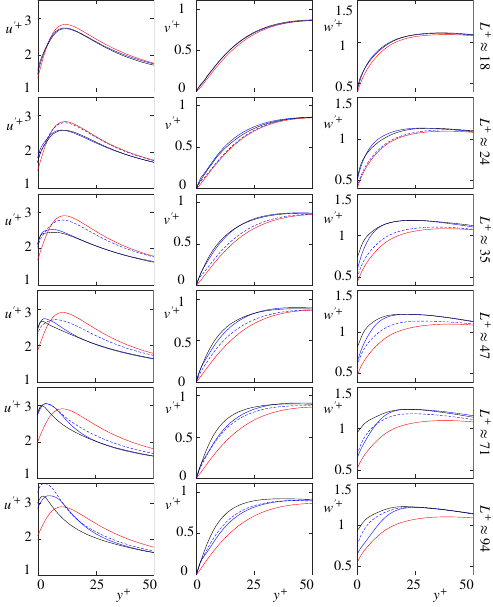}
\caption{R.m.s velocity fluctuations for collocated textures of sizes $L^+ \approx 20$--$100$ at $Re_{\tau} \approx 180$. \protect\blackline, texture-resolved simulations; \protect\redline slip-only simulations; \protect\bluedashedline, simulations with forcing based on conventional triple decomposition; \protect\blueline, simulations with forcing based on amplitude-modulated decomposition.}
\label{fig:stats_1}
\end{figure}

\begin{figure}
  \centering
  \vspace*{2mm}
  \includegraphics[trim=0 0 0 0,clip,width=0.7\textwidth]{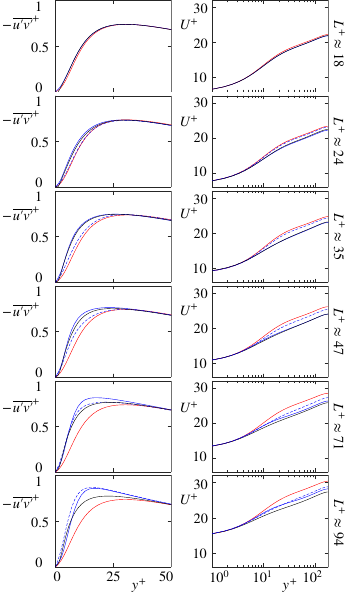}
\caption{Shear Reynolds stress and mean velocity profile for collocated textures of sizes $L^+ \approx 20$--$100$ at $Re_{\tau} \approx 180$. Line styles are as in figure \ref{fig:stats_1}.}
\label{fig:stats_2}
\end{figure}

The agreement exhibited by the roughness function between texture-resolved simulations
and texture-less simulations with amplitude-modulated forcing extends to 
other flow properties. This is the case for instance of one-point turbulent statistics. Figures \ref{fig:stats_1} and \ref{fig:stats_2}
portray the r.m.s. velocity fluctuations and the shear Reynolds stress and mean velocity profile for texture-resolved, slip-only and slip-plus-forcing
simulations, both using amplitude-modulated and conventional triple decomposition, across the range $L^+ \approx 20$--$100$. For $L^+ \approx 20$,
even slip-only simulations show good agreement with fully resolved ones.
This was reported by \cite{Fairhall19} as the limit size for which slip-lengths alone could capture the effect of the texture, as the texture-coherent flow is small in amplitude and confined to the immediate vicinity of the surface, and does therefore not alter the background 
turbulence significantly. The latter remains then smooth-wall like, other than by a shift in apparent origins
\citep{ibrahim2021smooth}. In agreement with this, adding forcing to model the effect of the texture-coherent fluctuations in this $L^+$ range has little effect and essentially does not alter the flow. 
As the texture size increases from $L^+ \approx 20$, though, the flow begins to depart from the smooth-wall-like behaviour that slip boundary conditions yield,
as shown in figures \ref{fig:stats_1} and \ref{fig:stats_2}, with a decrease
of the streamwise fluctuation intensity above $y^+ \approx 10$, and an increase throughout of the spanwise and wall-normal intensities and of the shear Reynolds stress. We note that
the latter is in all cases zero at the surface, a unique feature of slip-no slip textures
caused by the zero transpiration at $y=0$. In general,
the above modifications, which tend to decay sufficiently away from the wall, roughly at $y^+ \approx$ 50, are observed over slip/no-slip textures \citep{Seo15,Fairhall19} but also over rough surfaces \citep{orlandi2006,Abderrahaman-Elena19}. 

The addition of forcing using conventional triple decomposition can reproduce some of the departures from smooth-wall-like flow mentioned above for textured simulations, but forcing with amplitude modulation shows much better agreement with the resolved-texture cases. The agreement is excellent up to $L^+ \approx 50$ and first begins to break down for the spanwise velocity. This was
to be expected given the simplifications made in equation \ref{V2}, and has little effect on the Reynolds stress and, therefore, on the mean velocity profile and the drag. The agreement breaks down further for $L^+ \approx 70$, for which it begins to propagate into
$\Delta U ^+$, although it is still reasonable. For $L^+ \approx 100$ the departures become significant. We therefore identify this as the limit beyond which the model proposed here fails.

As mentioned in section \ref{sec:intro}, the differences in the Reynolds stress profile
shown in figure \ref{fig:stats_2}
are caused by the changes in the background turbulence. The latter are what the forcing
models ultimately aim to capture. The increased Reynolds stress causes a downward shift of the mean velocity profile away from the wall, and a corresponding reduction in $\Delta U^+$. This relationship can be quantified by integrating the streamwise momentum equation. Here we follow \cite{gomez2019turbulent}. A first integral gives
\begin{equation}
\frac{\d U^+}{\d y^+} + \tau_{uv}^+ = \frac{\delta'^+ - y^+}{\delta'^+},
\label{eq:mean_mome}
\end{equation}
where $\tau_{uv}$ is the shear Reynolds stress, including any dispersive stress, and $\delta'  = \delta + \ell_T$ is the effective half-height of the channel, which accounts for the background turbulence perceiving a virtual origin at $y=-\ell_T$. 
Integrating equation \ref{eq:mean_mome} once more in the wall-normal direction,
from the surface to a height $H$ sufficiently far above for all surface effects to have
vanished, gives
\begin{equation}
U^+ \left(H^+ \right) - U^+_\mathrm{slip}
+ \int_{0}^{H^+} \tau_{uv}^+\left(y^+ \right) \d y^+
= f,
\label{eq:mean_mome_integral}
\end{equation}
where $U_\mathrm{slip}=U(y^+=0)$ is the slip velocity, and $f$ is a simple function of $H^+$, $\delta'^+$ and $\ell_T^+$. The same integral
can be repeated for a reference smooth-wall flow at the same friction Reynolds number 
$\delta'^+$ between the corresponding heights, $y=\ell_T$ and $y=H+\ell_T$, yielding
\begin{equation}
U_\mathrm{S}^+ \left(H^+ + \ell_T^+\right) - U_\mathrm{S}^ + \left(\ell_T^+\right)
+ \int_{\ell_T^+}^{H^+ + \ell_T^+} \tau_{uv,\mathrm{S}}^+\left(y_S^+\right) \d y_S^+
= f,
\label{eq:mean_mom_smooth}
\end{equation}
where the subscript S denotes smooth-wall flow. The roughness function $\Delta U^+$
can then be obtained by subtracting equations \ref{eq:mean_mome_integral} and 
\ref{eq:mean_mom_smooth},
\begin{equation}
\begin{split}
\Delta U^+ &= U^+ \left(H^+ \right) - U_{S}^+ \left(H^+ + \ell_T^+\right)\\
& = U^+_\mathrm{slip} - U_\mathrm{S}^ + \left(\ell_T^+\right)
+ \int_{0}^{H^+} \!\!\left[ \tau_{uv}^+\left(y^+ \right)
- \tau_{uv,\mathrm{S}}^+ \left(y^+ + \ell_T^+ \right)\right] \d y^+.
\label{eq:du+_3}
\end{split}
\end{equation}

For slip-only simulations, turbulence is smooth-wall like and the integral in
equation \ref{eq:du+_3} is essentially zero \citep{Fairhall19,ibrahim2021smooth}. Near the wall,
$\d U^+/\d y^+\approx 1$, so for $\ell_x^+ \lesssim10$ we have $U^+_\mathrm{slip}
\approx\ell_x^+$. From equation \ref{eq:ellT+}, we have $\ell_T^+ \lesssim 4$, so also
$U_\mathrm{S}^ + \left(\ell_T^+\right)\approx \ell_T^+$. The roughness function reduces
then to the offset between the apparent origins for the mean flow and for turbulence,
$\Delta U^+ \approx \ell_x^+ - \ell_T^+$ \citep{Luchini96}.
For texture-resolved simulations, however, in addition to this origin offset, the increase in Reynolds stress causes further modifications to the mean velocity profile
and the roughness function. \citet{Fairhall19} observed that for the present slip/no-slip 
textures, with zero transpiration and dispersive stress at $y=0$, the contribution from the latter to $\tau_{uv}$ is negligible
up to at least texture sizes $L^+ \approx 50$, as shown in figure \ref{fig:dispersive}.
Those modifications would then be essentially caused by changes in the background, texture-incoherent 
turbulence alone. Figure \ref{fig:stats_2} shows that the forcing model with
amplitude modulation is able to capture the effect of these changes on
the Reynolds stress, and thus on $\Delta U^+$, while the one with conventional triple
decomposition captures those changes only partially.

\begin{figure}
  \centering
  \includegraphics[trim=0 0 0 0,clip,width=\textwidth]{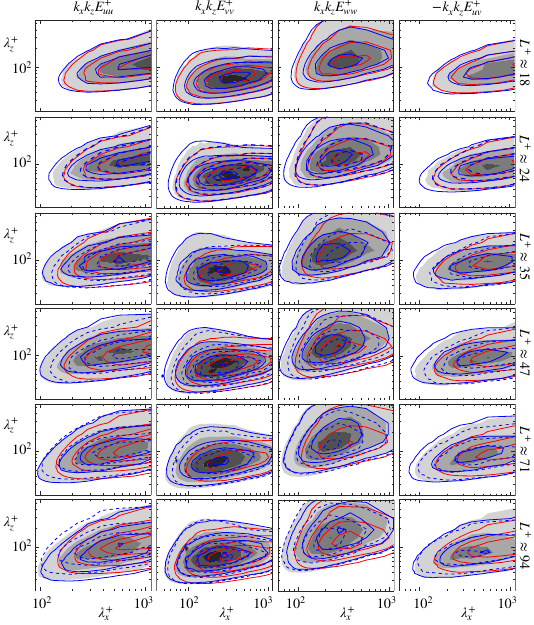}
\caption{Spectral energy densities of the velocity fluctuations and the shear Reynolds stress at $y^+ \approx 15$ for collocated textures of sizes $L^+ \approx 20$--$100$ at $Re_{\tau} \approx 180$. Shaded contours, texture-resolved simulations; \protect\redline slip-only simulations; \protect\bluedashedline, simulations with forcing based on conventional triple decomposition; \protect\blueline, simulations with forcing based on amplitude-modulated decomposition.}
\label{fig:all_spectra}
\end{figure}

The modifications in the background turbulence can be observed in more detail in the
spectral density maps of the different variables, as those portrayed in figure
\ref{fig:all_spectra}. Such maps show the contributions to the statistics shown
in figures \ref{fig:stats_1} and \ref{fig:stats_2} at a given height $y$ from different
streamwise and spanwise lengthscales. Figure
\ref{fig:all_spectra} displays the energy densities at
$y^+ \approx 15$, a height of intense r.m.s. fluctuations of the background
turbulence and also sufficiently above the surface for the texture-coherent flow
to be negligible. As observed by \cite{ibrahim2021smooth}, the signature of 
turbulence in slip-only simulations is essentially the same as over smooth 
walls. Compared to those cases, for fully resolved textures there is additional
energy in shorter streamwise scales, and also in wider spanwise scales
particularly for $v$. This effect is negligible for $L^+ \lesssim 20$,
but becomes increasingly marked for greater $L^+$.
Figure
\ref{fig:all_spectra} shows that slip-only models fail to capture this gradual change in the
dynamics of the background turbulence for $L^+ \gtrsim 25$. The forcing model based on conventional
triple decomposition is able to generate some additional energy in shorter
and wider scales, but not to the full extent observed in texture-resolved
simulations. In contrast, the forcing model based on amplitude modulation
can generate energy in all the necessary scales, and results in a good overall
collapse of the spectra with that for resolved textures up to $L^+ \approx 70$.
For larger $L^+$, deviations appear first in the core regions of the maps for $v$
and $w$ and in streamwise long scales of $u$ and $uv$. These spectra indicate again that
the additional forcing terms based on the amplitude-modulated decomposition
can capture the effect of the texture on the background turbulence, while
forcing based on conventional triple decomposition shows only a limited
improvement compared to slip-only simulations. 

%From the comparison of premultiplied energy spectra at height of 15 wall units above the turbulent virtual origin (figure \ref{fig:all_spectra}), where the signature of the texture is negligible, it shows that for $L^+ \approx 18$ the spectra of the turbulence over textures essentially collapse to those over slip lengths. This suggests that for the small textures size, the texture does not modify the dynamics of background turbulence. Hence the shift $\Delta U^+$ for the small texture size is well predicted by the homogeneous slip-length simulations. For the forcing model, the velocity fluctuations, mean velocity profile and premultiplied energy spectra show good agreements with those of the texture-resolving simulations as well as the conventional homogeneous-slip simulations. In this case, the non-linear interaction between the texture and background turbulence is weak. Hence the intensity of the additional forcing terms and therefore the improvement brought by introducing the forcing terms are small. 

\begin{figure}
  \centering
  \includegraphics[width=1.0\textwidth]{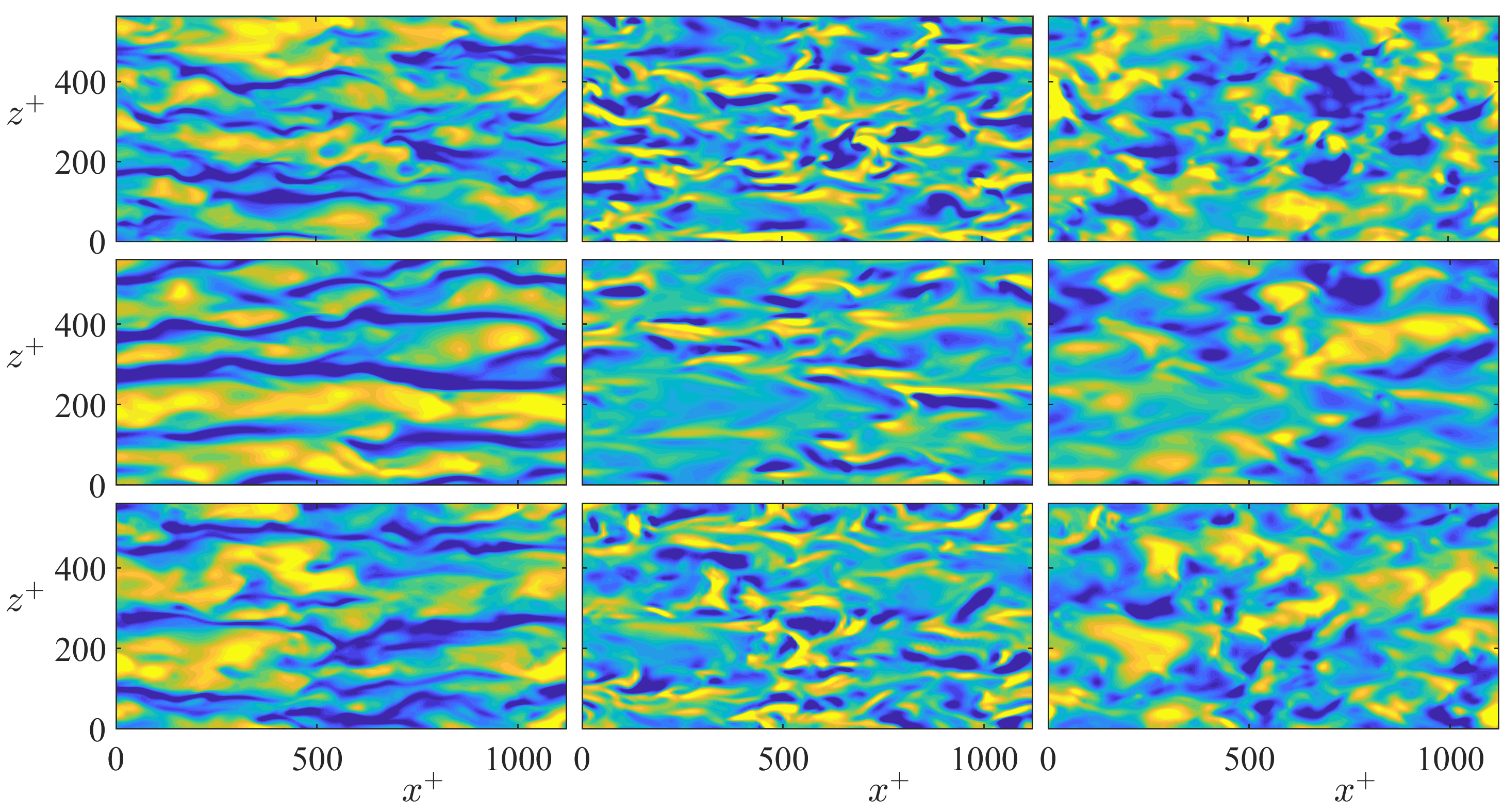}
\caption{Instantaneous realisations of the fluctuating $u$ (left), $v$ (centre) and $w$ (right) velocity components at $y^+ \approx 15$ for $L^+\approx 50$. Top row, texture-resolved simulation, TX47; middle row, slip-only simulation, SL47; bottom row, simulation with amplitude-modulated forcing, FA47. From blue to yellow, values are from -4.0 to 4.0 for $u$, from -1.2 to 1.2 for $v$, and from -2.2 to 2.2 for $w$, all in viscous units.}
\label{fig:velocity_field}
\end{figure}

To illustrate the general appearance of the flow, instantaneous realisations 
of the three velocity components for texture-resolved, slip-only, and
amplitude-modulated-forcing simulations are shown in figure \ref{fig:velocity_field}
at a height $y^+ \approx 15$ for texture size $L^+ \approx 50$. This size is large
enough to exhibit significant differences across models, yet not so large that
the forcing model begins to fail. While the slip-only simulation exhibits
canonical, smooth-wall-like-turbulence structures, both the texture-resolving
and the forcing-model simulations show a similar disruption of the latter, with
a reduction in the coherence of the streamwise velocity for scales of order
1000 wall units, as indicated also by the energy spectra. For the wall-normal
and spanwise velocities, the lengthscales of the turbulent eddies are also
shorter compared to those over homogeneous slip. 
At this $L^+ \approx 50$, the lengthscales and magnitude of the texture-coherent
flow near the wall become comparable to those of the background turbulence. The
spectra and instantaneous flow fields suggest a modification of the near-wall
dynamics through the disruption of streaks and quasi-streamwise vortices by
this texture-coherent flow. The similarity of the amplitude-modulated-forcing
flow field and the texture-resolved one suggests that the alteration of the
background turbulence by the presence of the texture can also be captured by
the forcing model.

Finally, to verify that the expected scaling in viscous units for textures whose effect is confined
to the vicinity of the wall \citep{Garcia-Mayoral2012} also holds for the corresponding models,
in addition to the simulations at $Re_{\tau} \approx 180$ we have conducted simulations at
$Re_{\tau} \approx 400$ for the collocated textures of size $L^+ \approx 35$ and
$L^+ \approx 100$. The results support this scaling,
and are presented and discussed in appendix \ref{appRe}.

%%%%%%%%%%%%%%%%%%%%%%%%%%%%%%%%%%%%%%%%%%%%%%%%%%%%%%%%%%%%%%%%%%%%%%%%%%%%%%%%
%%%%%%%%%%%%%%%%%%%%%%%%%%%%%%%%%%%%%%%%%%%%%%%%%%%%%%%%%%%%%%%%%%%%%%%%%%%%%%%%
\subsection{The limit of the forcing model}
\label{sec:limit}

The preceding discussion has shown how the model based on the amplitude-modulated
decomposition works well up to $L^+ \approx 70$, and how for $L^+ \approx 70$ and
$L^+ \approx 100$ deviations from the fully resolved simulations are increasingly
apparent. We have also mentioned that this is to be expected, as the assumption of
separation of scales between the two flow components ceases to hold. The decomposition
hinges on the idea that one component, the overlying background turbulence, excites the
other, the texture-coherent signature, as the response flow in the immediate vicinity
of the texture, which does not necessarily cease to hold once the lengthscales of the
two become comparable. The algebraic form, however, as laid out in equation
\ref{eq:u_decomposition} or more generally in equation \ref{eq:mod_decomp}, assumes
that the exciting overlying flow is on a much larger scale than the excited one,
in the spirit of \citet{Luchini91}. As $L^+$ increases, the breakdown of scale
separation occurs first for the cross flow components, as these are mainly produced
by the overlying quasi-streamwise vortices, with length and span of order $200\times20$
wall units, while the streamwise component is mainly produced by the streamwise
streaks, with length and span of order $1000\times100$ wall units \citep{Kline1967,Blackwelder1979,Smith1983}.

\begin{figure}
\vspace*{1mm}
  \centering
  \includegraphics[width=0.84\textwidth]{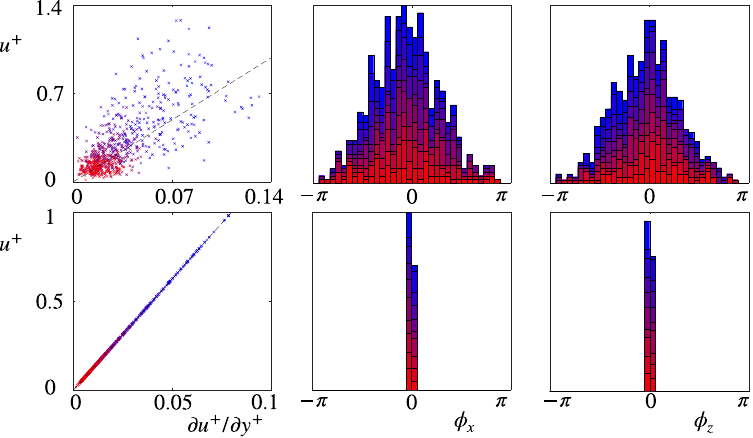}
\caption{Correlation over multiple instantaneous realisations of the streamwise
velocity $u$ and shear $\partial u/\partial y$ for texture size  $L^+ \approx 70$,
for wavelengths $\lambda_x^+\approx110$ to $1100$ and $\lambda_z^+\approx110$
to $550$ from red to blue. Left panels, magnitudes of $u$ and $\partial u/\partial y$;
centre and right panels, phases between the two in $x$ and $z$.
Top, full velocity signal; bottom, background turbulence component.
Dashed lines mark slip-lengths obtained from linear regression of the data displayed.}
\label{fig:breakd_u70}
%\end{figure}
%\begin{figure}
\vspace*{5mm}
  \centering
  \includegraphics[width=0.84\textwidth]{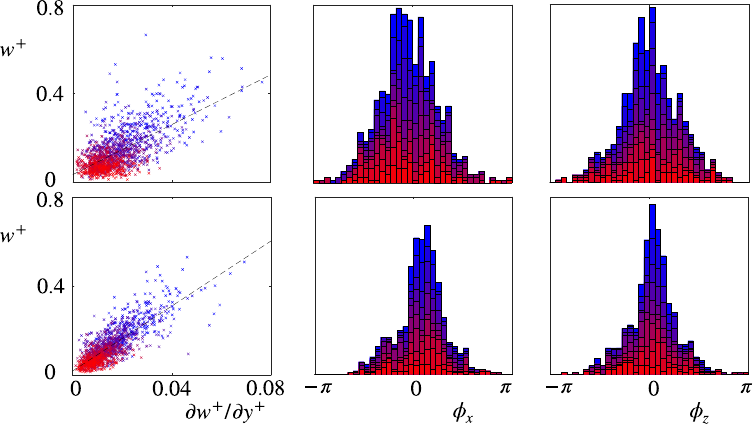}
\caption{Correlation over multiple instantaneous realisations of the streamwise
velocity $w$ and shear $\partial w/\partial y$ for texture size  $L^+ \approx 70$,
for wavelengths $\lambda_x^+\approx110$ to $1100$ and $\lambda_z^+\approx110$
to $550$ from red to blue. Left panels, magnitudes of $w$ and $\partial w/\partial y$;
centre and right panels, phases between the two in $x$ and $z$.
Top, full velocity signal; bottom, background turbulence component.
Dashed lines mark slip-lengths obtained from linear regression of the data displayed.}
\label{fig:breakd_w70}
\end{figure}

\begin{figure}
\vspace*{1mm}
  \centering
  \includegraphics[width=0.84\textwidth]{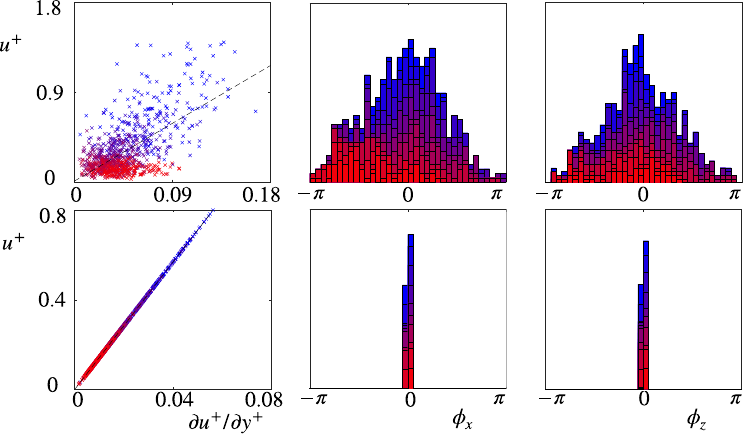}
\caption{Correlation over multiple instantaneous realisations of the streamwise
velocity $u$ and shear $\partial u/\partial y$ for texture size  $L^+ \approx 95$,
for wavelengths $\lambda_x^+\approx110$ to $1100$ and $\lambda_z^+\approx110$
to $550$ from red to blue. Left panels, magnitudes of $u$ and $\partial u/\partial y$;
centre and right panels, phases between the two in $x$ and $z$.
Top, full velocity signal; bottom, background turbulence component.
Dashed lines mark slip-lengths obtained from linear regression of the data displayed.}
\label{fig:breakd_u95}
%\end{figure}
%\begin{figure}
\vspace*{5mm}
  \centering
  \includegraphics[width=0.84\textwidth]{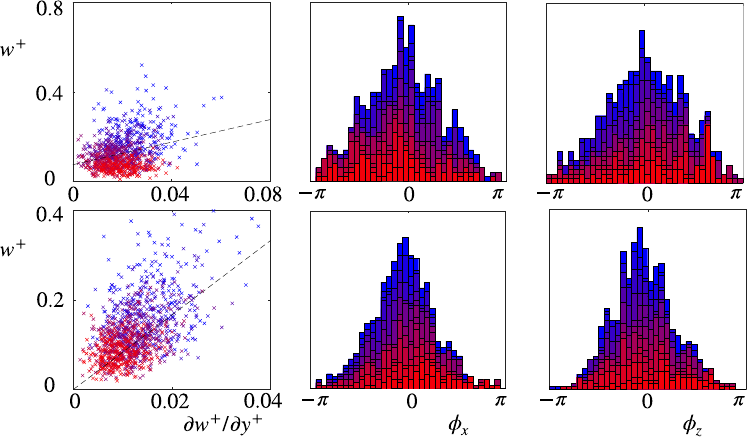}
\caption{Correlation over multiple instantaneous realisations of the streamwise
velocity $w$ and shear $\partial w/\partial y$ for texture size  $L^+ \approx 95$,
for wavelengths $\lambda_x^+\approx110$ to $1100$ and $\lambda_z^+\approx110$
to $550$ from red to blue. Left panels, magnitudes of $w$ and $\partial w/\partial y$;
centre and right panels, phases between the two in $x$ and $z$.
Top, full velocity signal; bottom, background turbulence component.
Dashed lines mark slip-lengths obtained from linear regression of the data displayed.}
\label{fig:breakd_w95}
\end{figure}

This difference can be observed
in the earlier loss of correlation at the wall between velocity and shear in the 
spanwise direction. \citet{Fairhall19} showed that the correlation held well for the
background turbulence up to
$L^+ \approx 50$ for both $u$ and $w$. Figures \ref{fig:breakd_u70} to
\ref{fig:breakd_w95} extend their analysis to the textures of size $L^+ \approx 70$
and $L^+ \approx 95$. The figures show that the decomposition using equation 
\ref{eq:mod_decomp} still results in an excellent recovery of the correlation between
$u$ and $\partial u/\partial y$ for both texture sizes,
both in terms of proportion and phase between them, which makes the value of $\ell_x$ estimated from this
data meaningful. The same is however not true for the correlation between $w$ and $\partial w/\partial y$,
which exhibits a considerable scatter both in terms of phase and magnitude. As a result, it is difficult to define a spanwise slip length $\ell_z$ meaningfully.
We note nevertheless that this is of minor importance with regards to setting the effective
boundary condition for $w$. At these large texture sizes, we have
$\ell_z^+\gtrsim7$, and thus the effect of the slip length in the boundary
condition for $w$ in equation \ref{eq:ellT+} is significantly
saturated, with variations in $\ell_z$ not resulting in
significant variations in $\ell_T$. In any event, the loss of correlation is
indicative of the background turbulence component obtained still being
contaminated by some amount of texture-induced signal,
and thus of the decomposition produced using equation 
\ref{eq:mod_decomp} breaking down.

\begin{figure}
  \centering
  \includegraphics[width=0.84\textwidth]{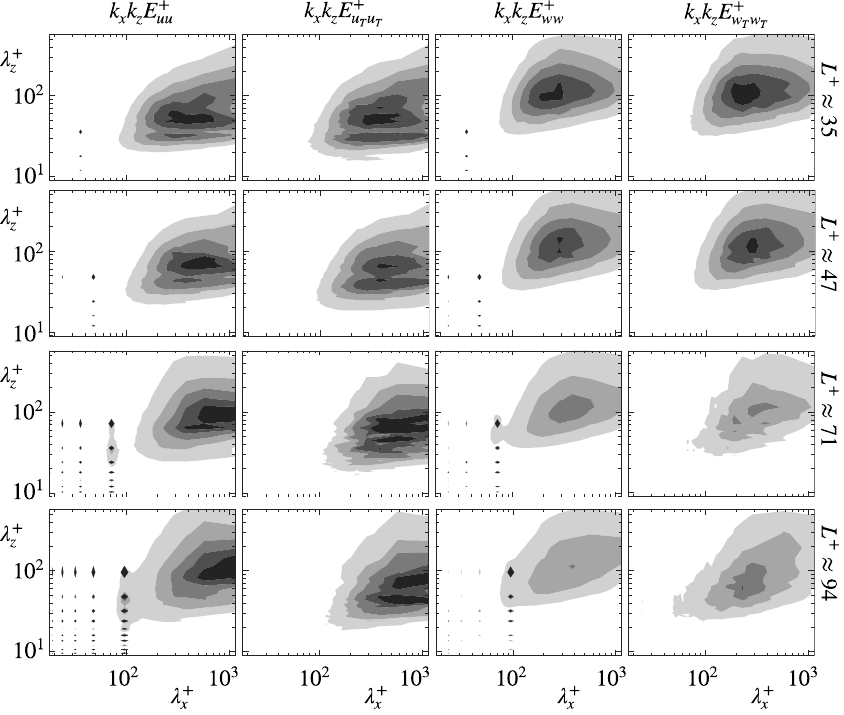}
\caption{Spectral energy densities at $y=0$ for the full streamwise and spanwise
velocity components, $u$ and $w$, and for the corresponding background
turbulence components as obtained using the amplitude-modulated decomposition
of equation \ref{eq:mod_decomp}, $u_T$ and $w_T$, for collocated square posts of sizes $L^+ \approx 35$,
47, 71 and 94.}
\label{fig:breakd_spec}
\end{figure}

This breakdown is perhaps even clearer in the spectral densities of the tangential
velocity components in the immediate vicinity of the
texture, both for the
background turbulence and the full texture-resolved
flow, as portrayed at $y=0$ in figure \ref{fig:breakd_spec}.
Up to texture size $L^+ \approx 50$ the main energetic region that can be observed in
the spectra of $u$ and $w$ is recovered in the spectra of the background-turbulence
components. However, for $L^+ \gtrsim 70$, the decomposition can successfully
filter out the contributions in the vicinity of the texture harmonics, which
is fully attributable to the texture-coherent flow and its modulation in amplitude
\citep{Abderrahaman-Elena19,Fairhall19}, but the spectral signature of the
background turbulence is visibly altered by the decomposition. This is
indicative of equation \ref{eq:mod_decomp} failing to produce the correct
signal for the background turbulence in the latter size range. Figure \ref{fig:breakd_spec} shows
that this happens once the lengthscales of the harmonics induced by the texture
begin to overlap significantly with the main spectral region of the background
turbulence, that is, once scale separation ceases to hold.

The results for texture sizes $L^+ \gtrsim 50$ portrayed in figures
\ref{fig:l_vs_delta_u} to \ref{fig:all_spectra} indicate that the deviations
from the texture-resolved results first occur for $w'$, and soon extend to the
other velocity components. The contribution of long structures of $u'$ and $uv'$
is particularly overpredicted, ultimately leading to an overprediction of the shear
Reynolds stress and an underprediction of the drag. We note
however that, even at $L^+ \approx 100$, the amplitude-modulated model performs
significantly better than the state-of-the-art slip-only models,
even if they fail to capture fully the dynamics at play in texture-resolved simulations.

The present break down of the model for $L^+ \gtrsim 70$ does not completely rule
out its applicability for larger textures, and rather limits its validity in the
present form of equations \ref{V2} and \ref{eq:bgd_2}. It is entirely possible that
a formulation of the flow decomposition not requiring scale separation could extend
its validity to larger sizes. The present formulation also neglects any texture-coherent
flow induced by cross-plane fluctuations. Such contributions to the texture-coherent
flow would become increasingly important for larger texture sizes, leading to additional
cross-advective forcing terms on the background turbulence. Their absence from the present formulation likely also contributes to its breakdown.
Dropping the above simplifications would lead to a significantly more
complex formulation, however, and is left for future work.

%%%%%%%%%%%%%%%%%%%%%%%%%%%%%%%%%%%%%%%%%%%%%%%%%%%%%%%%%%%%%%%%%%%%%%%%%%%%%%%%
%%%%%%%%%%%%%%%%%%%%%%%%%%%%%%%%%%%%%%%%%%%%%%%%%%%%%%%%%%%%%%%%%%%%%%%%%%%%%%%%
\subsection{Using a priori estimates for the texture-coherent flow}

The amplitude-modulated forcing model proposed in this paper shows good agreement with
texture-resolved simulations at least up to texture sizes $L^+ \approx 70$, both in terms
of drag prediction and turbulent statistics and structure. This strongly suggests that
the model has successfully identified the key physical mechanisms at play, i.e. the 
non-linear interactions between the texture-coherent flow and the background turbulence,
and is able to reproduce them without resolving the texture elements.
\rgm{Our central aim in the present work was to identify and understand this mechanism, but ultimately
it would also be interesting to use this understanding  to design predictive tools. The results presented thus far do} not have a truly predictive character, as they require \rgm{a posteriori} information on the
texture-coherent flow obtained by ensemble averaging of texture-resolved simulations. In order to be fully 
predictive, the model would \rgm{instead} need a priori estimates of the texture-coherent flow. 
\rgm{Such estimates are readily available, and their use is discussed below. The results are good, but we note, however, that this
does not add to the understanding of the physical mechanism, and its usefulness as a predictive tool is somewhat limited, as it would still require running DNS-resolution simulations, even if without the texture-related resolution requirements. This is nevertheless a useful first step for the development of future models that circumvent the need for DNS altogether.}

As mentioned in \S \ref{sec:intro}, \citet{Abderrahaman-Elena19} observed that the ensemble
average flow over small textures resembles the flow induced by a steady homogeneous
overlying shear at matching $L^+$, which can be obtained from laminar steady simulations
of a single periodic unit of texture. This is in essence an extension of the Stokes-flow
computations used by \citet{Luchini91}, \citet{Kamrin2010} and \citet{Luchini2013} to
predict protrusion heights or slip lengths, with the addition of inertial
terms. The laminar computations of \citet{Abderrahaman-Elena19} were extended for the
slip/no-slip textures considered in this paper in \citet{adams2021laminar}. In the
latter, a synthetic eddy-viscosity obtained from smooth-wall DNS was added to represent
the effect of the background turbulence on the shape of the mean velocity profile under
a driving pressure gradient. \rgm{This eddy viscosity was essentially used so that the laminar computations yielded mean profiles more realistic than Couette or Poiseuille ones, although this only had a marginal effect on the texture-coherent flow, in its weaker regions away from the surface. It is analogue to a \citet{VanDriest1956}  or Cess \citep[see][]{Reynolds1967} conventional model, but was used instead because the latter was found to produce a poor surrogate for the smooth-wall profile at our low $Re_{\tau} \approx 180$.} Compared to the texture-resolved DNSs of \citet{Fairhall19},
the resulting flow fields for $L^+\approx35$ overpredicted the coherent $\widetilde{u}_u$ by roughly 20\%,
and showed good agreement in $\widetilde{v}_u$ and $\widetilde{w}_u$, yielding overall
a reasonable estimate for the texture-coherent flow.

\begin{figure}
  \centering
  \vspace*{1mm}
  \includegraphics[width=.96\textwidth]{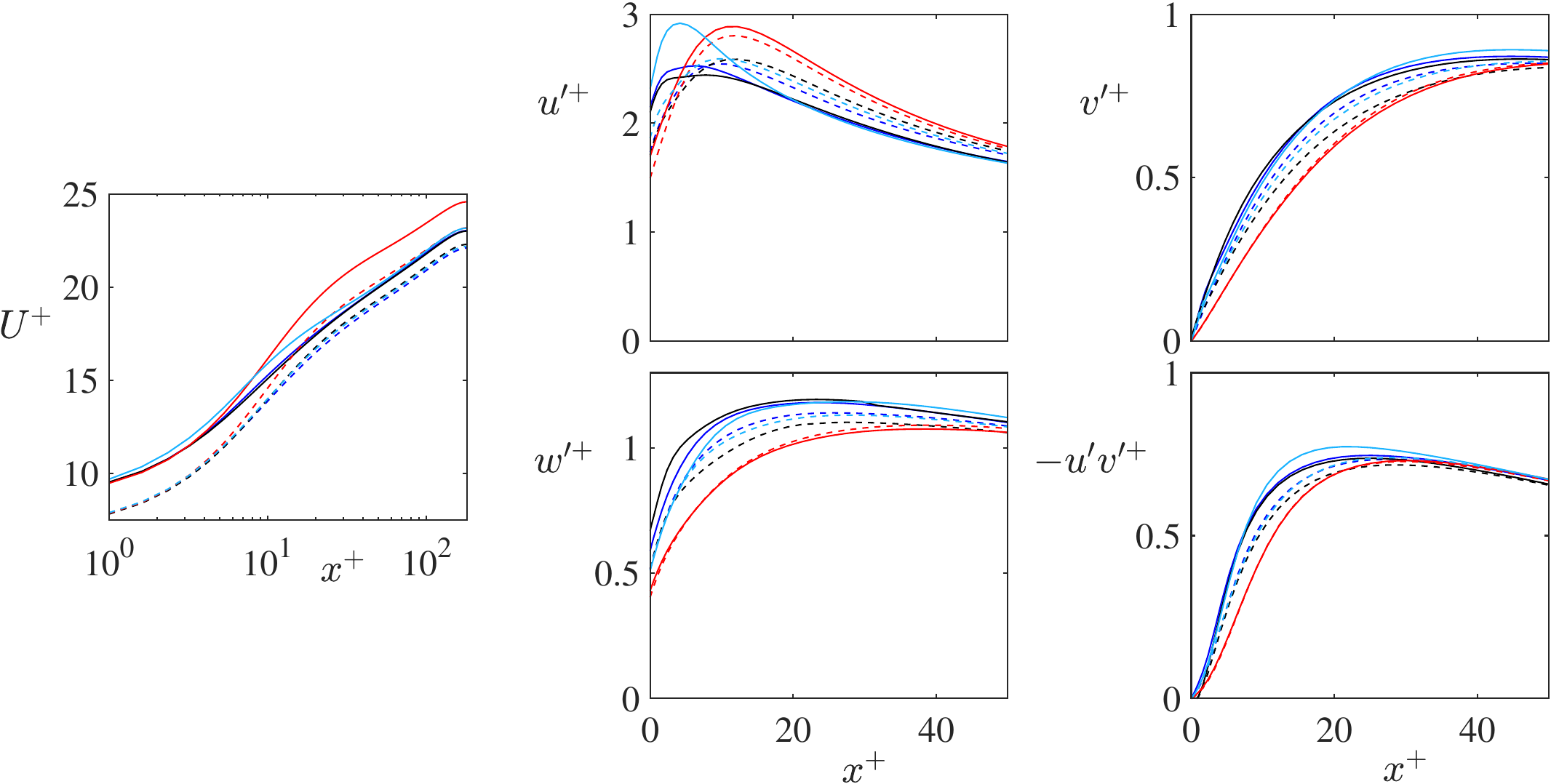}
  \vspace*{-1mm}
\caption{Mean velocity profile, r.m.s velocity fluctuations and Reynolds shear stress for simulations with $L^+ \approx 24$ (dashed lines) and $L^+ \approx 35$ (solid lines) at $Re_{\tau} \approx 180$. \protect\blackline,~texture-resolved simulations, TX24 and TX35; \protect\redline, slip-only simulations, SL24 and SL35; \protect\blueline, simulations with amplitude-modulated forcing, FA24 and FA35; \protect\lightblueline, simulations with amplitude-modulated forcing using an a priori surrogate for the texture-coherent flow, FA24S and FA35S.}
\label{fig:laminar_stats}
\end{figure}

\begin{figure}
  \centering
  \includegraphics[trim=0 0 0 0,clip,width=\textwidth]{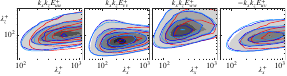}
  \vspace*{-5mm}
\caption{ Spectral energy densities of the velocity fluctuations and the Reynolds shear stress at $y^+ \approx 15$ for $L^+ \approx 35$ at $Re_{\tau} \approx 180$. Filled contours, texture-resolved simulation, TX35; \protect\redline, slip-only simulation, SL35; \protect\blueline, simulation with amplitude-modulated forcing, FA35; \protect\lightblueline, simulation with amplitude-modulated forcing using an a priori surrogate for the texture-coherent flow, FA35S.}
\label{fig:laminar_spectra}
\end{figure}

As a first step to explore the predictive capabilities of the amplitude-modulated forcing
model, we study its performance when the texture-coherent flow from DNSs is replaced by the
a priori surrogates of \citet{adams2021laminar} for $L^+\approx24$ and $L^+\approx35$. Results are shown in figures \ref{fig:laminar_stats} and \ref{fig:laminar_spectra}.
Figure \ref{fig:laminar_stats} shows excellent agreement in the r.m.s. velocity fluctuations,
the Reynolds shear stress and $\Delta U^+$, except for an overshoot in $u'$ for the larger texture $L^+\approx35$ in the immediate
vicinity of the wall, $y^+\lesssim15$. This is likely caused by the excessive intensity
in $\widetilde{u}_u$ in the surrogate texture-coherent flow, as mentioned above. The
overshoot does nevertheless not propagate into the Reynolds stress or any of the other
velocity components, and is also confined to a narrow region near the texture, so it
essentially does not alter the drag or the near-wall dynamics and dissipation, mainly
governed by cross-plane velocities induced by bursts and quasi-streamwise vortices
\citep{ibrahim2021smooth,Khorasani2022,jimenez2022}. Similar results can also be observed in the spectral
densities, shown for the three velocity components and the Reynolds shear stress at
$y^+\approx15$ in figure \ref{fig:laminar_spectra}
for $L^+\approx35$. The surrogate texture-coherent flow
is able to produce an energetic signature in the same short wavelengths exhibited by
texture-resolved simulations and those with forcing based on the a posteriori
texture-coherent flow, and not present for smooth-wall or slip-only simulations.
Although there is room for improvement, especially in the prediction of the near-wall
peak in $u'$, these results suggest that the need to conduct simulations that resolve
the texture can be fully circumvented, and that it is possible to resolve the background turbulence, and predict the drag and other flow properties using
only a priori data on the texture detail, significantly reducing computational costs.

%illustrate the comparison of r.m.s. fluctuations and spectral energy densities of the velocity components and the Reynolds stress for TX35, SL35, FA35 and the forcing simulation based on the modified laminar flow. The results obtained from the forcing model based on modified laminar flow is much more accurate compared to the conventional homogeneous-slip simulations, especially for the Reynolds stress and mean velocity profiles. This shows the potential of \textit{a priori} predicting the dynamics of background turbulence over textures. 

%%%%%%%%%%%%%%%%%%%%%%%%%%%%%%%%%%%%%%%%%%%%%%%%%%%%%%%%%%%%%%%%%%%%%%%%%%%%%%%%
%%%%%%%%%%%%%%%%%%%%%%%%%%%%%%%%%%%%%%%%%%%%%%%%%%%%%%%%%%%%%%%%%%%%%%%%%%%%%%%%
%%%%%%%%%%%%%%%%%%%%%%%%%%%%%%%%%%%%%%%%%%%%%%%%%%%%%%%%%%%%%%%%%%%%%%%%%%%%%%%%
%%%%%%%%%%%%%%%%%%%%%%%%%%%%%%%%%%%%%%%%%%%%%%%%%%%%%%%%%%%%%%%%%%%%%%%%%%%%%%%%
\section{Conclusions}
\label{sec:con}

The simulation and prediction of turbulence over textured surfaces requires a spatial
resolution that can be more demanding than that needed to resolve the turbulence itself,
and can become prohibitive for texture sizes in the range $L^+\approx5$-$100$. This is
a relevant range for applications, and includes the working range for drag-reducing
textures and the transitionally rough range for drag-increasing ones, beyond which
the effect on drag usually asymptotes. In this paper we have investigated the effect
of the texture detailed geometry on the background turbulence, with the \rgm{ultimate} aim to sidestep the
need to resolve that geometry directly. As a first approach, we have focused on a particularly
simple, idealised texture, consisting of a slip/no-slip pattern on an otherwise flat
surface, a popular model for superhydrophobic surfaces. Such slip/no-slip textures
have some unique properties that simplify their analysis, namely that the effective
boundary conditions for the overlying turbulence are well known and characterised,
reducing to slip, Robin conditions for the tangential velocities and zero transpiration;
and that the wall-normal velocity and Reynolds stress are zero at the reference,
interfacial plane, and remain negligible in its vicinity. This allowed \citet{Fairhall19}
to identify that drag degraded due to additional Reynolds stresses occurring above
the surface, rather than arising at the surface and propagating into the flow above.
They proposed that such Reynolds stresses had to originate from the
non-linear interaction of the texture-coherent flow with the background turbulence.
This interaction became significant and could not be neglected for $L^+\gtrsim25$, and would need
to be accounted for in any model in addition to the effective slip boundary conditions,
which still held up in this range of sizes and at least up to $L^+\approx50$, 

Following up from \citet{Fairhall19}, we have analysed how the decomposition of
the flow into a texture-coherent and a background-turbulence components propagates into
the governing equations of the two components separately. We have first decomposed
the flow using conventional triple decomposition, which produces a much simpler
set of governing equations. However, as observed by \citet{Abderrahaman-Elena19},
this decomposition results in a residual texture-coherent signature in the background turbulence.
To separate fully the texture-coherent signal and obtain a texture-incoherent background
turbulence, \citet{Abderrahaman-Elena19} proposed a decomposition in which the latter
modulates in amplitude the former. Here we have analysed the governing equations
for the background turbulence that result from this decomposition. For both the conventional
and the amplitude-modulated triple decomposition, we have argued that the background turbulence is governed by Navier-Stokes equations with additional,
forcing terms, arising from the cross advective terms between both flow components.

To assess the above hypothesis, we have conducted a series of simulations in which
the background turbulence was fully resolved, but the texture was replaced by the corresponding
effective boundary conditions plus the forcing terms in the momentum equations mentioned
above. For the formulation based on conventional triple decomposition, the results
are partially improved compared to using effective boundary conditions alone.
For the formulation based on the amplitude-modulated decomposition, the results
show excellent agreement up to texture sizes $L^+\approx50$, and begin to deviate for
$L^+\approx70$. This is the case both for the prediction of drag and $\Delta U^+$, and
for the statistical properties of turbulence, including r.m.s. velocity fluctuations,
Reynolds stresses and mean velocity profiles. In particular, the addition of forcing
terms can accurately generate fluctuations at shorter wavelengths than smooth-wall flows,
which are present for geometry-resolved simulations but which the effective boundary
conditions are unable to generate on their own.
Once deviations begin to occur, for texture sizes $L^+\gtrsim70$, they first manifest
in the spanwise velocity component. This is also the component for which the flow
decomposition first begins to break down, likely due to the span of background-turbulence
eddies becoming comparable to the texture size earlier than their length, which disrupts
the implicit assumption of separation of scales embedded in the amplitude-modulated
decomposition. Concurrently, the texture-coherent motions excited not only by
the overlying streamwise velocity, but by the cross velocity components, cease
to be negligible, and therefore our simplifications in the forcing also cease to hold.
 
The above results strongly suggest that the key effects of surface texture on the overlying
turbulence are imposing a set of effective boundary conditions but, critically, also altering
the momentum equations through the non-linear interaction with the texture-coherent flow in
the advective terms. While this is useful
in terms of identifying physical mechanisms, it falls short in terms of predictive power,
as quantifying that interaction requires prior knowledge of the texture-coherent flow, but
the latter only becomes available from the postprocessing of fully resolved flows. We have
therefore also explored the possibility of using a priori surrogates for the
texture-coherent flow, based on the laminar steady flow about a single periodic unit of
texture. Preliminary results show good agreement, motivating further work.

The present study has only considered a particularly simple type of surface texture. The
results are encouraging, but would need to be further tested in more complex and
general textures, mainly roughness. Many of the simplifications possible here, e.g.
the effective boundary conditions reducing to homogeneous slip and zero transpiration, or
accounting only for the texture-coherent flow induced by the overlying streamwise
velocity, will likely not apply in the general case, calling for a more complete and complex framework.

\backsection[Acknowledgements]
{Computational resources were provided by the University of Cambridge Research Computing Service under EPSRC Tier-2 grant
EP/P020259/1, and by the UK 'ARCHER2' system under PRACE (DECI-17) project pr1u1702 and
EPSRC project e776.
For the purpose of open access, the authors have applied a Creative Commons Attribution
(CC BY) licence to any Author Accepted Manuscript version arising from this submission.}

\backsection[Declaration of interests]{The authors report no conflict of interest.}

\backsection[Author ORCIDs]{\\
Chris Fairhall, https://orcid.org/0000-0002-2529-3650;\\
R. Garc{\'i}a-Mayoral, https://orcid.org/0000-0001-5572-2607.}
%%%%%%%%%%%%%%%%%%%%%%%%%%%%%%%%%%%%%%%%%%%%%%%%%%%%%%%%%%%%%%%%%%%%%%%%%%%%%%%%
%%%%%%%%%%%%%%%%%%%%%%%%%%%%%%%%%%%%%%%%%%%%%%%%%%%%%%%%%%%%%%%%%%%%%%%%%%%%%%%%
%%%%%%%%%%%%%%%%%%%%%%%%%%%%%%%%%%%%%%%%%%%%%%%%%%%%%%%%%%%%%%%%%%%%%%%%%%%%%%%%
%%%%%%%%%%%%%%%%%%%%%%%%%%%%%%%%%%%%%%%%%%%%%%%%%%%%%%%%%%%%%%%%%%%%%%%%%%%%%%%%
%\clearpage
\appendix

\section{Magnitude of the forcing term and the residual in the momentum equations}
\label{appResid}

\begin{figure}
  \centering
  \includegraphics[trim=0 0 0  0,clip,width=\textwidth]{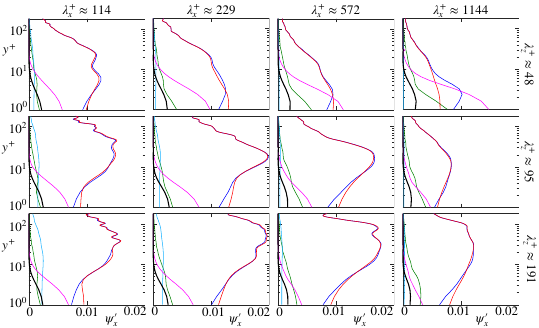}
\caption{Time-r.m.s intensity $\psi_x'$ of the terms in the streamwise
component of the momentum equation derived for the background turbulence assuming
amplitude-modulated texture-coherent flow, equation \ref{eq:bgd_2}, for the case
of collocated texture with $L^+\approx35$. \protect\blueline, advective term;
\protect\greenline, viscous term; \protect\lightblueline, pressure-gradient term; \protect\redline,
temporal-derivative term; \protect\magentaline, non-linear forcing term $\boldsymbol{N_b'}$; \protect\thicblackline, residual $\boldsymbol{R}$.}
\label{fig:residual_1}
\end{figure}

\begin{figure}
  \centering
      \includegraphics[trim=0 0 0 0,clip,width=\textwidth]{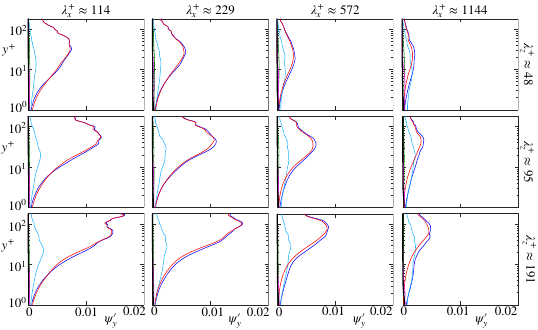}
\caption{Time-r.m.s intensity $\psi_y'$ of the terms in the wall-normal
component of the momentum equation derived for the background turbulence assuming
amplitude-modulated texture-coherent flow, equation \ref{eq:bgd_2}, for the case
of collocated texture with $L^+\approx35$. \protect\blueline, advective term;
\protect\greenline, viscous term; \protect\lightblueline, pressure-gradient term; \protect\redline,
temporal-derivative term; \protect\magentaline, non-linear forcing term $\boldsymbol{N_b'}$; \protect\thicblackline, residual $\boldsymbol{R}$.}
\label{fig:residual_2}
\end{figure}

\begin{figure}
  \centering
      \includegraphics[trim=0 0 0 0,clip,width=\textwidth]{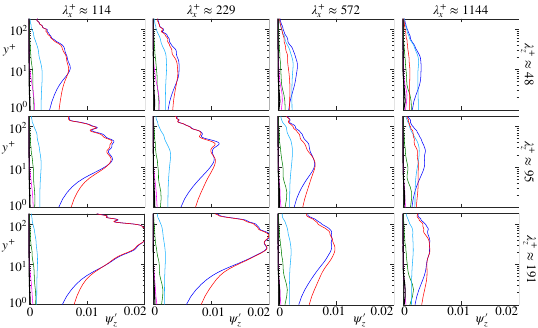}
\caption{Time-r.m.s intensity $\psi_z'$ of the terms in the spanwise
component of the momentum equation derived for the background turbulence assuming
amplitude-modulated texture-coherent flow, equation \ref{eq:bgd_2}, for the case
of collocated texture with $L^+\approx35$. \protect\blueline, advective term;
\protect\greenline, viscous term; \protect\lightblueline, pressure-gradient term; \protect\redline,
temporal-derivative term; \protect\magentaline, non-linear forcing term $\boldsymbol{N_b'}$; \protect\thicblackline, residual $\boldsymbol{R}$.}
\label{fig:residual_3}
\end{figure}

This appendix portrays evidence of the relative importance of the forcing term
$\boldsymbol{N_b'}$ and the residual $\boldsymbol{R}$ in equation \ref{eq:bgd_2}.
The residual is neglected in the governing equations of the simulations presented
in \S \ref{sec:res} with forcing based on the amplitude-modulated decomposition,
and it therefore needs to be small for the proposed model to hold.
Conversely, if the magnitude of $\boldsymbol{N_b'}$ was small, we would expect that
retaining this term had little effect on the dynamics of the background turbulence.

Figures \ref{fig:residual_1} to \ref{fig:residual_3} portray
the time-r.m.s. magnitude of the different terms in equation \ref{eq:bgd_2} from the
texture-resolved simulation of collocated posts with texture size $L^+\approx35$
(case TX35). Figures \ref{fig:residual_1}, \ref{fig:residual_2} and \ref{fig:residual_3}
portray results for the streamwise, wall-normal and spanwise components of the momentum
equation respectively, for a representative set of $x$-$z$ wavenumbers and as a function
of height $y$. The figures show that the dynamics are mostly dominated by the interplay
between temporal and advective terms, but in the vicinity of the wall, for $y^+ \lesssim 10$,
the forcing term, which is neglected in slip-only simulations, can become comparable to,
and even larger that, the latter two. This is particularly the case for the streamwise momentum
equation, as illustrated by figure \ref{fig:residual_1}, and across the
whole range of wavelengths. For the spanwise and wall-normal momentum equations,
the explicit contribution of $\boldsymbol{N_b'}$ appears to be less significant. 
In turn, the residual is also negligible in the spanwise and wall-normal momentum
equations. It is typically a fraction of $\boldsymbol{N_b'}$ of order 1/4-1/3 in the streamwise one,
and thus reasonably smaller than the leading-order terms. These results back up neglecting
$\boldsymbol{R}$ when implementing the amplitude-modulated forcing model in \S \ref{sec:res}.

%%%%%%%%%%%%%%%%%%%%%%%%%%%%%%%%%%%%%%%%%%%%%%%%%%%%%%%%%%%%%%%%%%%%%%%%%%%%%%%%
%%%%%%%%%%%%%%%%%%%%%%%%%%%%%%%%%%%%%%%%%%%%%%%%%%%%%%%%%%%%%%%%%%%%%%%%%%%%%%%%
%\clearpage
\section{Simulations for staggered-posts textures}
\label{appStagg}

\begin{figure}
  \centering
\vspace*{-1mm}
\includegraphics[width=.8\textwidth]{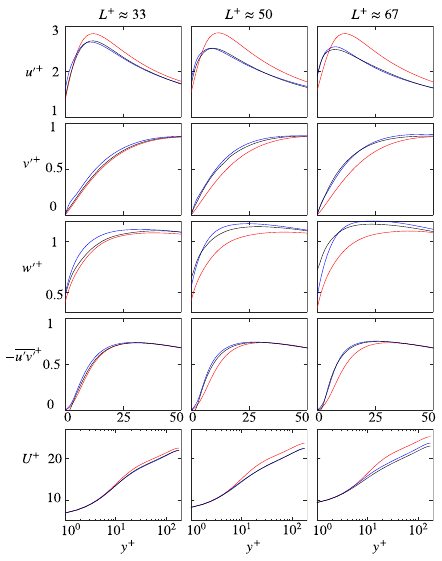}\hspace*{12mm}
\vspace*{-1mm}
\caption{R.m.s. velocity fluctuations, shear Reynolds stress and mean velocity profile for staggered-posts textures with $L^+\approx35$-$70$. \protect\blackline, texture-resolved simulations; \protect\redline,~slip-only simulations; \protect\blueline, simulations with forcing based on amplitude-modulated decomposition.}
\label{fig:stats_staggered}
\end{figure}

\begin{figure}
  \centering
  \vspace*{2mm}
  \includegraphics[trim=0 0 0  0,clip,width=\textwidth]{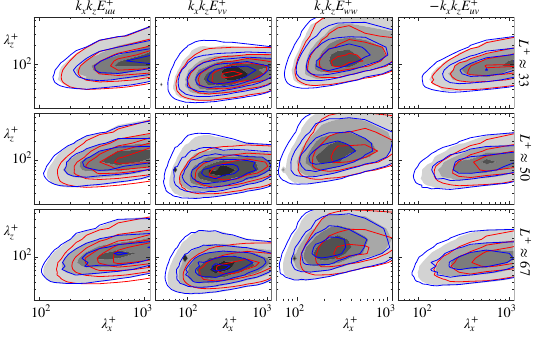}
  \vspace*{-1mm}
\caption{Spectral energy densities of the velocity fluctuations and the Reynolds shear stress at $y^+ \approx 15$ for staggered texture arrangements. Shaded contours, texture-resolved simulations; \protect\redline, slip-only simulations;  \protect\blueline, simulations with forcing based on amplitude-modulated decomposition.}
\label{fig:all_spectra_stg}
\end{figure}

Figures \ref{fig:stats_staggered} and \ref{fig:all_spectra_stg} portray results from the 
simulations for staggered-posts configurations, as sketched in figure \ref{fig:text_arr}.
The aim of these simulations was to verify if the forcing model using amplitude-modulated
decomposition, first tested for a collocated-posts texture, would also be
applicable to other surface arrangements. In additions to simulations fully resolving
the texture and simulations with forcing, we have also conducted simulations implementing slip
boundary conditions alone, to serve as a benchmark for comparison. These simulations complement and expand the data set of \citet{Fairhall19}.

We have conducted simulations for texture sizes $L^+\approx35$, $50$ and $70$, aiming to span the size range
for which slip-only simulations begin to deviate from texture-resolved ones, yet the
texture size is not large enough for the decomposition and the forcing model to break down. The simulation parameters and resulting roughness functions are listed in table \ref{tab:simulations}.

The results follow the same trends of the collocated-textures cases. Figure
\ref{fig:stats_staggered} shows that the r.m.s. velocity fluctuations for texture-resolved
simulations exhibit differences with the slip-only ones, which display smooth-wall-like turbulence 
\citep{Fairhall19,ibrahim2021smooth}. These differences are small for $L^+\approx35$, 
but become increasingly pronounced for larger texture sizes. Their most salient features
are a decrease in near-wall peak $u'$ and an increase in $v'$ and $w'$. These differences
eventually result in a significant increase in the shear Reynolds stress, and a corresponding
downward shift of the mean velocity profile away from the wall. The forcing model is able to accurately
capture these differences, and produces results very close to those of texture-resolved
simulations, with small deviations first appearing in $w'$, as discussed in \S \ref{sec:res} for collocated textures.

The spectral densities portrayed in figure \ref{fig:all_spectra_stg} are also
consistent with the observations in \S \ref{sec:res} for collocated textures. Compared to
smooth-wall turbulence, in texture-resolving simulations have additional
energy particularly in shorter scales in the streamwise direction. Slip-only simulations
are not able to produce this modification in turbulence, and their spectrum remains
essentially smooth wall like as discussed in \citet{Fairhall19} and \citet{ibrahim2021smooth}. In contrast, the forcing model is
able to produce the spectral energy densities of texture-resolved simulations with excellent agreement.

%%%%%%%%%%%%%%%%%%%%%%%%%%%%%%%%%%%%%%%%%%%%%%%%%%%%%%%%%%%%%%%%%%%%%%%%%%%%%%%%
%%%%%%%%%%%%%%%%%%%%%%%%%%%%%%%%%%%%%%%%%%%%%%%%%%%%%%%%%%%%%%%%%%%%%%%%%%%%%%%%
%\clearpage
\section{Effect of Reynolds number}
\label{appRe}

\begin{figure}
  \centering
      \includegraphics[width=.72\textwidth]{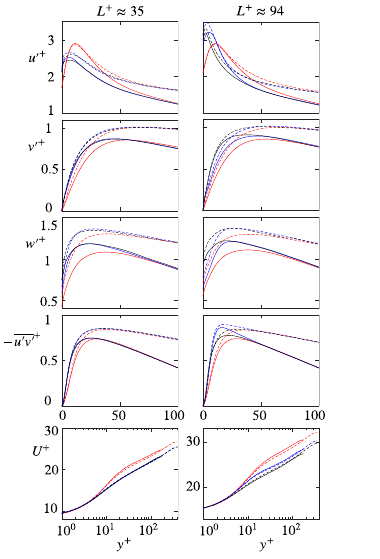}
%\vspace*{-1mm}
\caption{R.m.s. velocity fluctuations, shear Reynolds stress and mean velocity profile for
simulations at $Re_{\tau} \approx 180$ (solid) and $Re_{\tau} \approx 400$ (dashed). \protect\blackline,
TX35; \protect\blackdashedline, TX35H; \protect\redline, SL35; 
\protect\reddashedline, SL35H;
\protect\blueline, FA35; \protect\bluedashedline, FA35H.}
\label{fig:re_comp_stats}
\end{figure}

\begin{figure}
  \centering
      \includegraphics[trim=0 0 0  0,clip,width=\textwidth]{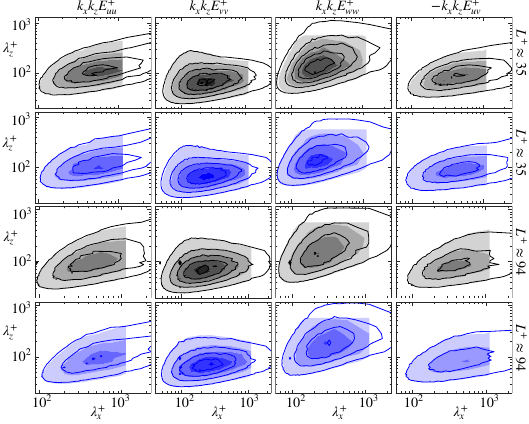}
\caption{Spectral energy densities of the velocity fluctuations and the Reynolds shear stress at $y^+ \approx 15$ for $L^+ \approx 35$ and $L^+ \approx 100$ at different $Re_{\tau}$. Grey filled contours, texture-resolved simulations at $Re_{\tau} \approx 180$; \protect\blackline, at $Re_{\tau} \approx 400$. Blue filled contours, simulations with forcing based on amplitude-modulated decomposition at $Re_{\tau} \approx 180$; \protect\blueline, at $Re_{\tau} \approx 400$.}
\label{fig:re_comp_spectra}
\end{figure}

The main set of simulations discussed in this paper were conducted at $Re_{\tau} \approx 180$.
To verify that our observations would scale out to other Reynolds numbers, we have conducted
a reduced set of simulations at $Re_{\tau} \approx 400$, otherwise matching in inner scaling setups from the
main set, with results shown in figures \ref{fig:re_comp_stats} and \ref{fig:re_comp_spectra}.
We have chosen the collocated layouts with $L^+ \approx 35$ and $L^+ \approx 100$, as
one texture size for which the texture-resolved and the forcing simulations show good
agreement between them, but differences with the slip-only simulation, and one texture size for which the forcing model shows signs of breaking down.

The results are consistent with our usual observations for other surface topologies
\citep{Garcia-Mayoral2012,Fairhall19,sharma2020sparse,sharma2020dense,Hao2024}. Near the wall,
the same effects are produced in inner units for matching simulations, with results
eventually collapsing to smooth wall data at equal $Re_{\tau}$ for $y^+\gtrsim50$. Differences
in this region are observable, but entirely attributable to the variations in $Re_{\tau}$, as
they are the same observed for smooth-wall flows \citep{Moser1999}. These are evidenced for
instance in the higher peak levels of the r..m.s. velocity fluctuations and the shear
Reynolds stress near $y^+\gtrsim25$, as shown in figure \ref{fig:re_comp_stats}. The
spectral densities portrayed in \ref{fig:re_comp_spectra} confirm these results. The spectral
energy content at each wavelength matches when scaled in inner units for corresponding simulations,
with larger-scale content only resolved for the higher $Re_{\tau} \approx 400$, as made
possible by the larger simulation domain.
Overall, the results suggest that the effect of the model, like the effect of the texture, scales in inner units as expected, and can be
extrapolated to and used at different Reynolds numbers.

\bibliographystyle{jfm}
\bibliography{wenxiong_refs}

\end{document}